\documentclass[aps,prd,showpacs,floatfix,notitlepage,a4paper]{revtex4}
\usepackage{amsfonts,amsmath,units,wasysym,epsfig,graphicx,verbatim,color,subfigure,graphicx,bm,mathrsfs,lipsum,hyperref}

\usepackage[utf8]{inputenc}
\usepackage[normalem]{ulem}  

\begin{document}

\newcommand{\IUCAA}{Inter-University Centre for Astronomy and
  Astrophysics, Post Bag 4, Ganeshkhind, Pune 411 007, India}

\newcommand{\WSU}{Department of Physics \& Astronomy, Washington State University, 1245 Webster, Pullman, WA 99164-2814, U.S.A}

\title{The tidal deformability of an anisotropic compact star: Implications of GW170817}

\author{Bhaskar Biswas $^{\rm 1}$}
\email{bhaskarb@iucaa.in}
\author{Sukanta Bose $^{\rm 1, 2}$}
\affiliation{$^{\rm 1}$ \IUCAA,$^{\rm 2}$\WSU}

\begin{abstract}
 We use gravitational wave (GW) and electromagnetic (EM) observations of GW170817 to constrain the extent of pressure anisotropy in it. 
While it is quite likely that the pressure inside a neutron star is mostly isotropic, certain physical processes or characteristics, such as phase transitions in nuclear matter or the presence of strong magnetic fields, can introduce 
pressure anisotropy. In this work, we show that anisotropic pressure in  neutron stars can reduce their tidal deformability  substantially. For the anisotropy-pressure model of Bowers and Liang and a couple of relativistic EOSs -- DDH$\delta$ and GM1 -- we demonstrate that this reduction in spherical neutron stars with masses in the range of 1 to 2 $M_\odot$ can be 23\% to 46\%. This suggests that certain EOSs that are ruled out by GW170817 observations, under assumptions of pressure isotropy, can become viable if the stars had a significant enough anisotropic pressure component, but do not violate causality.  We also show how the inference of the star radius can be used to rule out certain EOSs (such as GM1), even for high anisotropic pressure, because their radii are larger than what the observations find. 

\end{abstract}

\preprint{LIGO-DCC-P1800396}

\maketitle

\section{Introduction}
The recent detection of gravitational waves (GWs) from the binary neutron star (BNS) merger event GW170817~\cite{Abbott2017} has initiated a new way to probe and constrain the equation of state of compact stars \cite{Abbott:2018exr,Radice2018}. In the inspiral stage of the binary coalescence the tidal deformation of the orbiting stars leaves an imprint on the emitted GW signal~\cite{Flanagan2008,Hinderer2008,Binnington2009,Damour2009}. This imprint carries information about the composition of the star. Unfortunately, properties of neutron star matter at very high density are not fully understood. Therefore, modelling the star requires one to make certain assumptions about its interior. 

One of the most common assumptions made in studies of the equilibrium structure of a neutron star is that its pressure is isotropic. Specifically, in a spherically symmetric neutron star, the radial pressure and the transverse pressure are taken to be equal. Interestingly, it has been argued in other studies that this equality may not always hold; in other words, the pressure in a neutron star can have an anisotropic component.
In basic terms, pressure anisotropy can arise whenever the velocity distribution of particles in a fluid is anisotropic, which in turn can owe its origin to the presence of magnetic fields, turbulence, convection, etc.~\cite{Herrera}. 
There are several studies (see, e.g., Refs.~\cite{Kippenhan2012,Ruderman1972,Canuto1974}) that suggest that at very high densities relativistic interactions between nucleons can make the pressure anisotropic. In the density range of $0.2 $ fm$^{-3}$ to $1$ fm$^{-3}$, superdense nuclear matter makes a phase transition to almost equal numbers of protons, neutrons and $\pi^-$ particles \cite{Sawer1972}. The $\pi^-$ particles condense to a plane wave state of momentum that can be as large as $\approx$ 170 Mev/c. This condensation causes a drastic reduction in pressure, which softens the equation of state along the radial direction ~\cite{Sawer1974}.

Another interesting scenario arises owing to strong magnetic fields that neutron stars are known to possess. Indeed, many neutron stars have magnetic fields with strength $10^{12}-10^{13}$~G; and there is evidence for the existence of supermagnetized neutron stars with magnetic fields as large as $10^{14}-10^{15}$~G~\cite{magnetar1, magnetar2}. Magnetic pressure associated with such strong magnetic fields can also induce pressure anisotropy.

Presence of P type superfluid or solid core can also introduce pressure anisotropy \cite{Kippenhan2012}, where interactions among the P type superfluid nucleons produce the anisotropy. It has been shown by Herrera et al. \cite{Herrera} that use of the two-fluid model naturally predicts pressure anisotropy. An example of the two-fluid model is Superfluid Helium II, in the context of the Landau theory~\cite{Landau}.

Finally, it is also known that in certain braneworld models of gravity, with an extra spatial dimension, the corrections induced in Einstein's equations on the four-dimensional brane can be modeled as a stress energy tensor with anisotropic pressure~\cite{Kabirda}. 

For the aforementioned reasons we explore here how gravitational wave observations of binary neutron stars and, in the process, measurements of their macroscopic parameters, such as their mass, radius and the tidal deformability parameter \cite{Flanagan2008,Hinderer2008},
can be used to test the presence or absence of pressure anisotropy 
in these stars. There exists a large body of work on equilibrium configuration and oscillations of anisotropic neutron stars~\cite{Bowers and Liang,HH,CHEW,S1,B,KBD,MM,Bondi,Chan,GM,PM,HPOF,HN,HM,Papakostas01,MDH,BI,Doneva2012,Silva,Yagi1,Yagi2,yagi3,jose2016,Raposo2018}. These studies suggest that anisotropy in pressure, if present with a non-negligible magnitude, can have a significant effect on the mass-radius relationship and, therefore, the compactness of the star. That in turn leads one to enquire what effect, if any, anisotropy may have on the GWs emitted during a binary coalescence. 

Yagi and Yunes~\cite{Yagi1} came close to addressing this matter when they compared the tidal deformability of slowly rotating neutron stars in the presence and absence of pressure anisotropy. Their work was mainly focused on how much the anisotropy affects the universal relation between moment of inertia, tidal Love number and quadrupole moment. In this present work we calculate the tidal deformability~\cite{Hinderer2008} of a static anisotropic compact star whose background is taken to be spherically symmetric. We also use a different EOS for pressure anisoptropy, namely, the one pioneered by Bowers and Liang~\cite{Bowers and Liang}.
We show that there are regions in that EOS parameter space that give rise to unphysical  stellar configurations (owing to the existence of regions where causality would be violated). After discarding such configurations from further study, 
we calculate the change in the tidal deformability parameter of neutron stars for a few cases of anisotropic pressure EOS, in an otherwise standard relativistic equation of state (EOS). We find that the presence of pressure anisotropy generally reduces its tidal deformability, for a fixed stellar mass. We demonstrate how this property allows certain relativistic EOSs, for a range of pressure anisotropy magnitudes, to remain viable in light of GW170817. We also use that observation and universality relations between the tidal deformability parameter and stellar compactness, deduced here, to constrain the pressure anisotropy parameter. Finally, we explain how future observations of GWs from binary neutron stars can tighten this constrain further.

Throughout this paper, we set the gravitational constant $G$ and the speed of light in vacuum $c$ to unity, except when computing observational quantities, such as the second Love number or the tidal deformability parameter, for comparison with observations.

\section{Equilibrium configurations of anisotropic compact stars}
\label{sec2}
 \begin{figure*}
\begin{center}
\begin{tabular}{cc}
\includegraphics[width=0.45\textwidth]{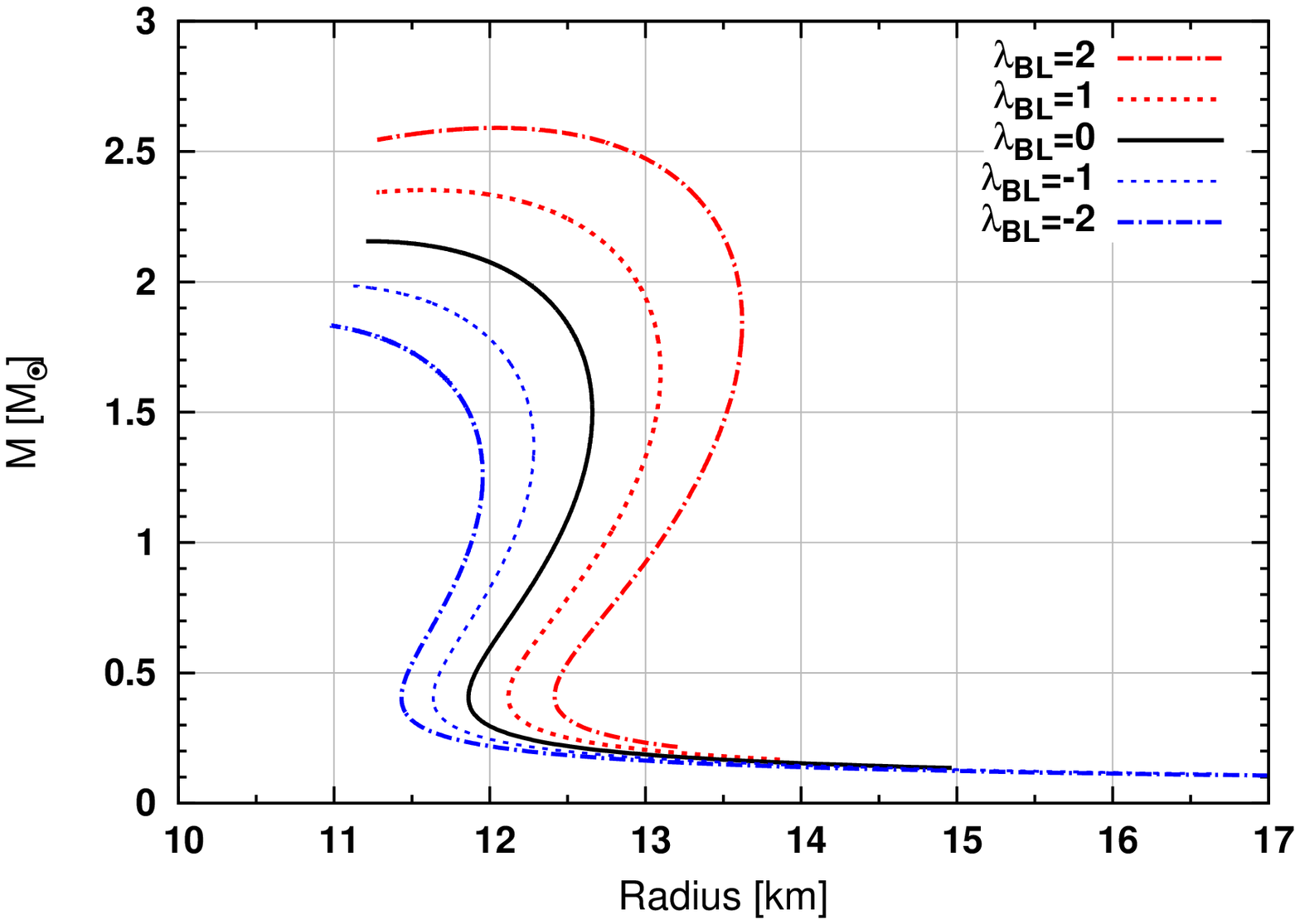}&
\includegraphics[width=0.45\textwidth]{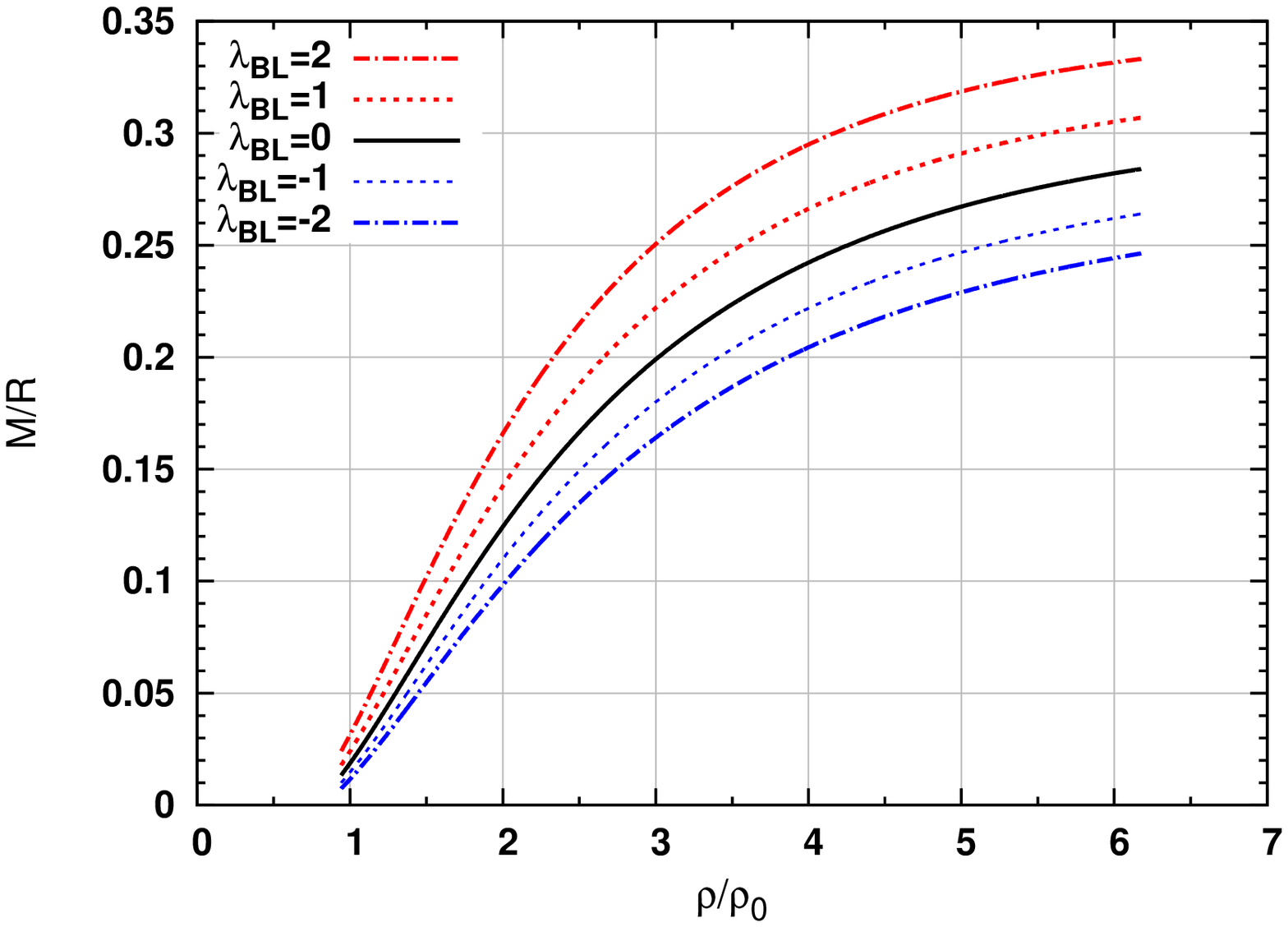}\\
\includegraphics[width=0.45\textwidth]{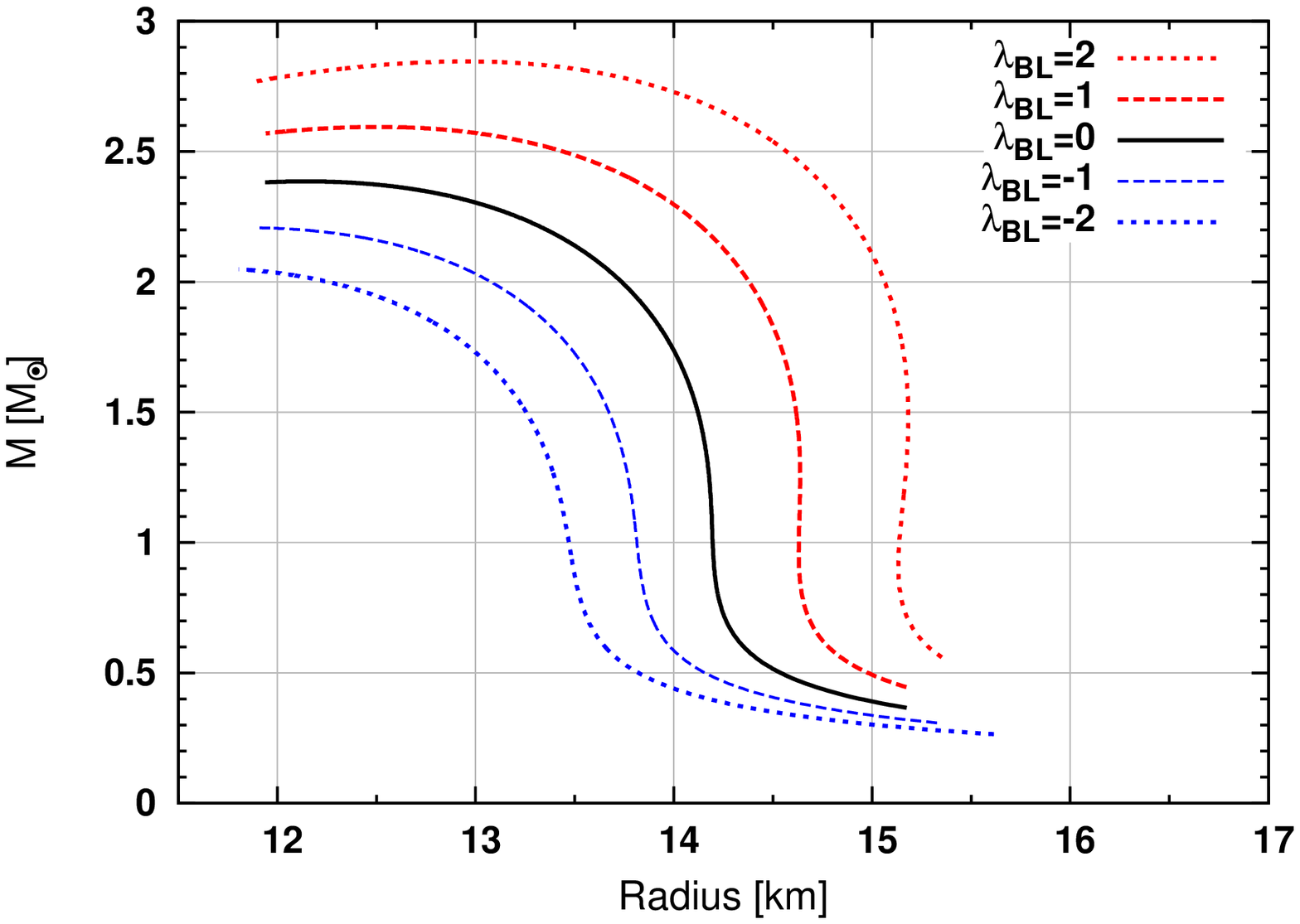}&
\includegraphics[width=0.45\textwidth]{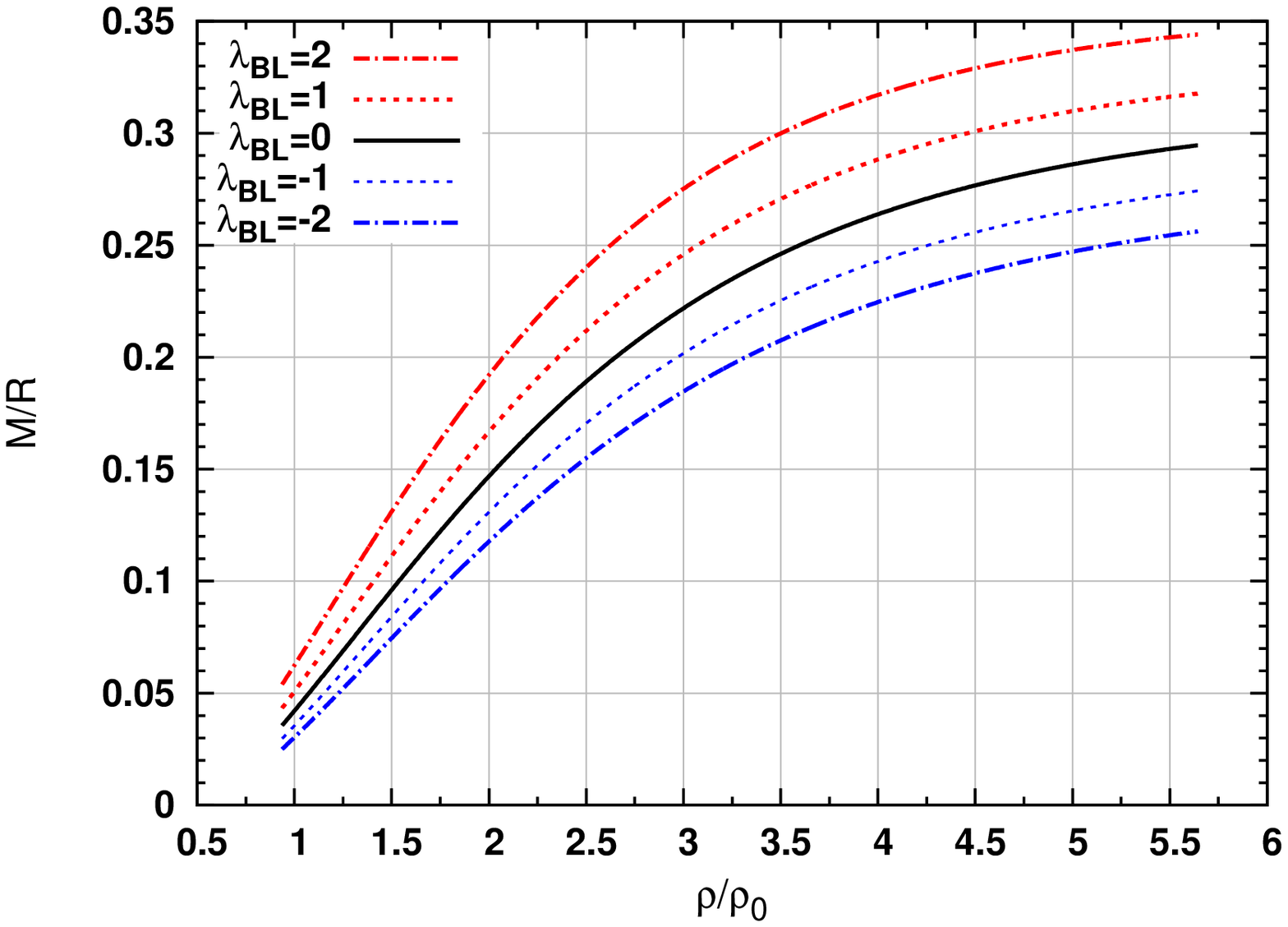}\\
\end{tabular}
\end{center}
\caption{The mass-radius relationship (left panel) and compactness of the neutron star as a function of normalized baryon density (right panel)  for several values of the anisotropic parameter $\lambda_{BL}$ using EOSs DDH$\delta$ (top panel) and GM1 (bottom panel).}
\label{fig:stellar_properties}
\end{figure*}

\begin{figure*}[ht]
\begin{center}
\begin{tabular}{cc}
\includegraphics[width=0.45\textwidth]{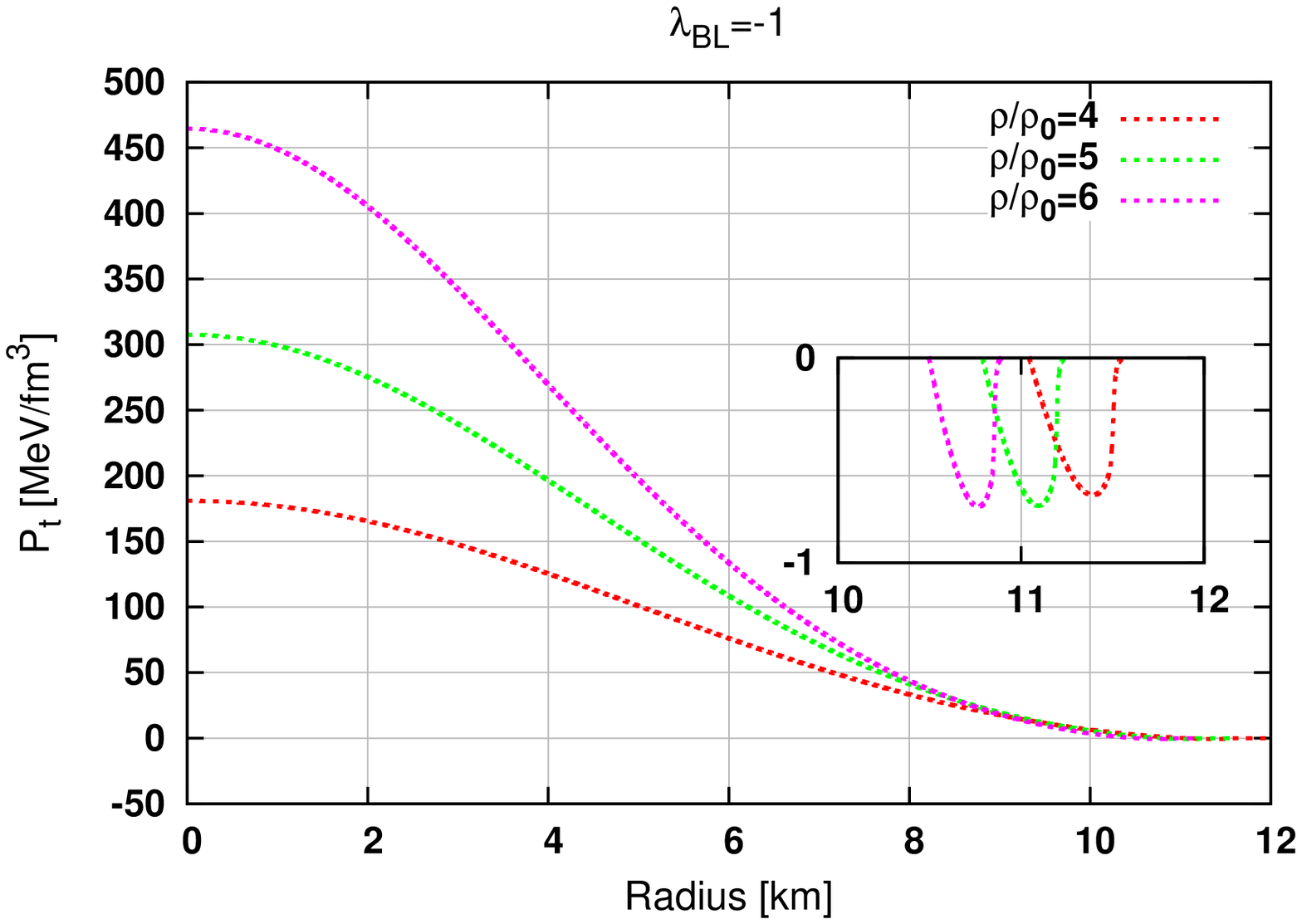}&
\includegraphics[width=0.45\textwidth]{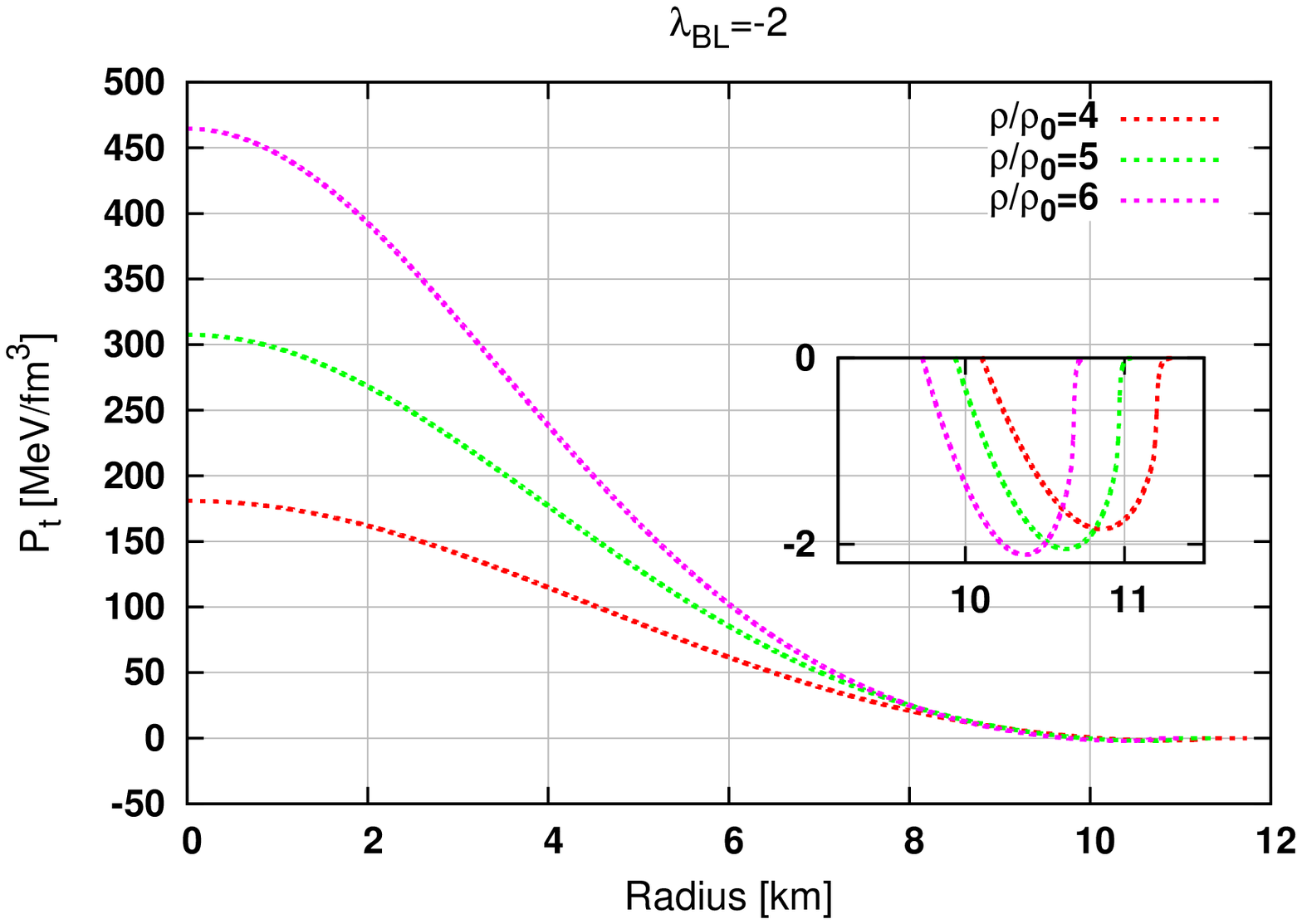}\\
\includegraphics[width=0.45\textwidth]{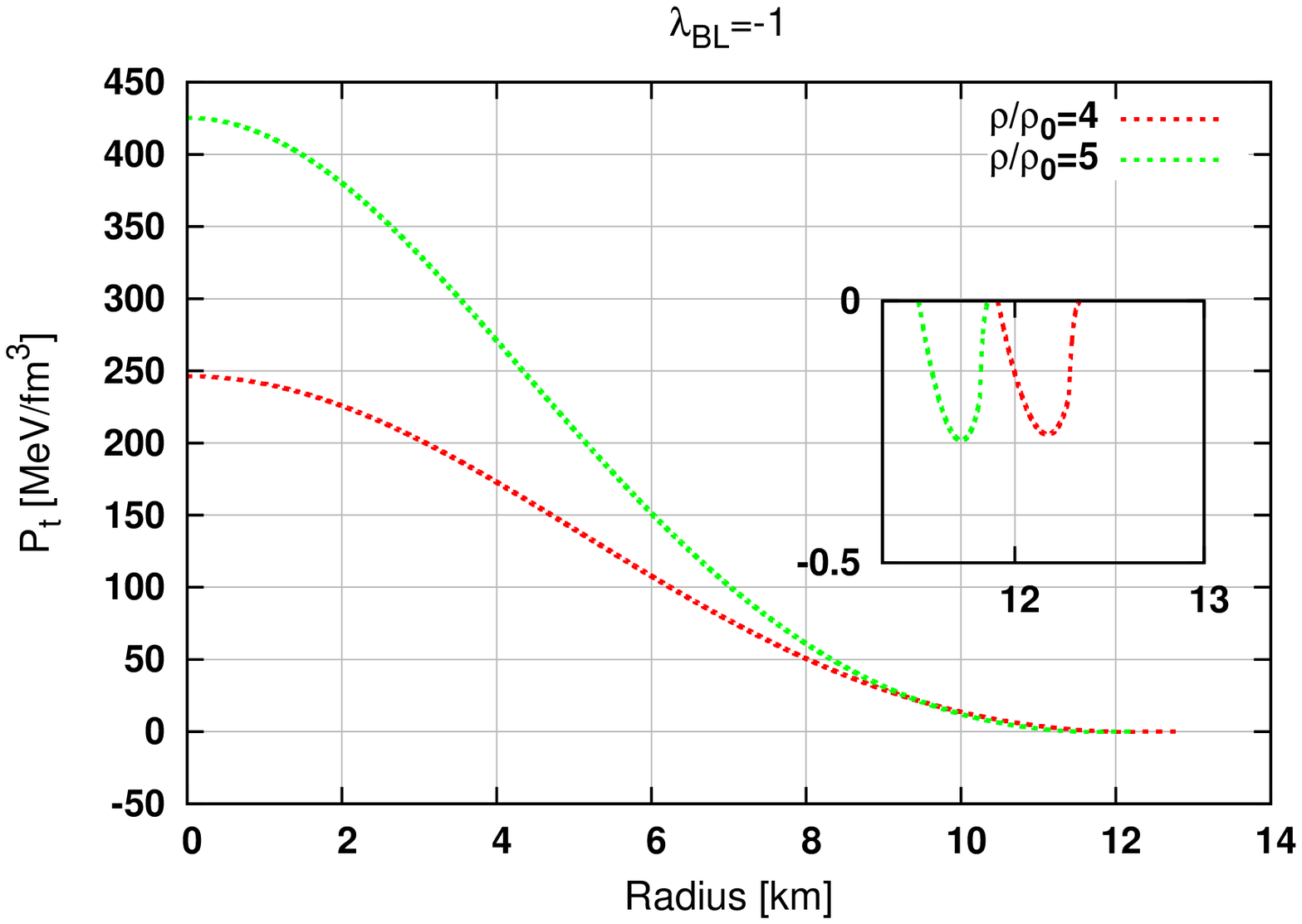}&
\includegraphics[width=0.45\textwidth]{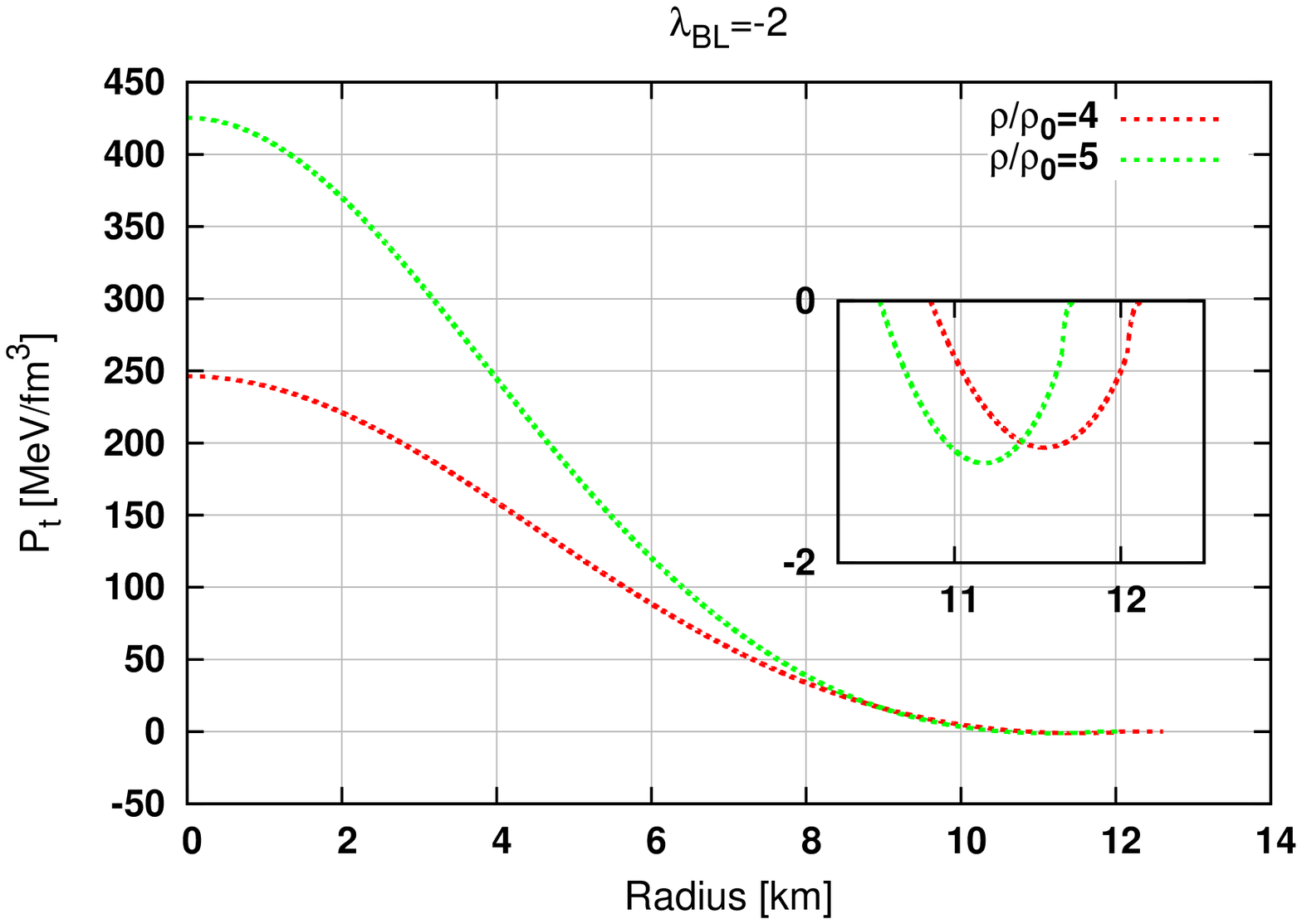}\\
\end{tabular}
\end{center}
\caption{Radial profile of transverse pressure for different values of central density using DDH$\delta$ (top panel) and GM1 (bottom panel) EOS: $\lambda_{BL}=-1$ (left panel) and $\lambda_{BL}=-2$ (right panel).}
\label{fig:transverse_pressure}
\end{figure*}
We consider a static, spherically symmetric fluid distribution with an anisotropic component. Its stress-energy tensor is given as:

\begin{equation}
    T_{\nu}^{\mu}=diag (\rho,-p_r,-p_t,-p_t)\,,
\end{equation}
where $\rho$ is the density, and the transverse pressure $p_t$ differs from the usual radial pressure $p_r$ owing to  anisotropy. In Schwarzschild coordinates the metric 
takes the form
\begin{equation}
    ds^2=g_{\alpha \beta}^{(0)}dx^{\alpha}dx^{\beta}= e^{\nu}dt^2-e^{\lambda}dr^2-r^2d\theta^2-r^2 \sin^2\theta  d\varphi^2\,.
\end{equation}
Using this matter distribution and spacetime geometry in Einstein's equations gives one the following modified Tolman-Oppenheimer-Volkov (TOV) equations,
\begin{equation}
 \frac{dp_r}{dr}=-\frac{\left(\rho + p_r \right)\left(m + 4\pi r^3 p_r \right)}{r\left(r -2m\right)} +\frac{2}{r} (p_t -p_r)\,,
  \label{tov1:eps}
\end{equation}
\begin{equation}
\frac{dm}{dr}=4\pi r^{2}{\rho}\,,
    \label{tov2:eps}
\end{equation}
where $m(r)$ is the mass enclosed within areal radius $r$.

To close this system of equations one considers two separate EOSs for $p_r$ and $p_t$. We will assume barotropic EOS for radial pressure, $p_r=p_r(\rho)$. Specifically, we study stars with two different EOSs based on the relativistic mean field (RMF) parametrization, namely, DDH$\delta$ \cite{GNPA_2004} and GM1 \cite{GMPR_1991} in beta equilibrium. In both cases, for the crust an EOS by Douchin and Haensel \cite{DHA_2001} is added below a density of $10^{-3}$ fm$^{-3}$. For transverse pressure, we consider the functional form given by Bowers and Liang~\cite{Bowers and Liang},
\begin{equation}
\label{Anisotropy_eos}
p_t = p + \frac{1}{3} \lambda_{BL} \frac{\rho +3p}{1-2m/r}(\rho +p)r^2 \,,
\end{equation}
where the constant $\lambda_{BL}$ is a measure of anisotropy.
Note that in Eq.~(\ref{Anisotropy_eos}), and hereafter, we use $p \equiv p_r$ to denote the radial pressure. For this particular choice of anisotropic EOS, the pressure anisotropy $(p_t - p)$ (which affects the second term on the right-hand side of Eq.~(\ref{tov1:eps})) must vanish quadratically with $r$ at the center of the star in order to yield regular stellar solutions. This form of $p_t$ was also motivated in Ref.~\cite{Bowers and Liang} by the consideration that at least a part of the anisotropy is gravitationally induced, thereby, giving rise to its nonlinear dependence on $p$. The boundary condition $p(r=R)=0$ determines the radius $R$ of the star. For all physically acceptable solutions we must have $p,~p_t\geq 0$ inside the star. 

Following Silva et al.~\cite{Silva} we begin by examining solutions in the relatively narrow range $-2\leq \lambda_{BL} \leq 2$ around isotropy, which is when $\lambda_{BL} = 0$.
For DDH$\delta$ and GM1, we find that when the transverse pressure is higher than the radial pressure (i.e., $\lambda_{BL} >0$) the star can support more mass against gravitational collapse compared to the opposite situation (i.e., $\lambda_{BL} <0$). We also find that, for a fixed central density, compactness of the star increases (decreases) if the transverse pressure exceeds (falls below) radial pressure. These properties are depicted in Fig.~\ref{fig:stellar_properties}. Is it possible to observationally constrain the degree of anisotropy in a neutron star? Below we present a way to do so with gravitational wave observations. 

Focusing first on {\em negative} values of $\lambda_{BL}$, we find some evidence that the transverse pressure in such configurations may not always be positive. (See Fig.~\ref{fig:transverse_pressure} for $\lambda_{BL}=-1,-2$.) Since these may correspond to unphysical solutions, we choose to study them in a separate work. However, for smaller negative values of $\lambda_{BL}$, the condition $p_t \geq 0$ can be respected everywhere in the star. Nevertheless, in those cases the anisotropic effects will be smaller; 
we do not study such cases here. Below we exclusively study the positive $\lambda_{BL}$ solutions, with particular attention on how their tidal deformability may differ from the corresponding $\lambda_{BL} = 0$ solutions.

\begin{figure*}[ht]
\begin{center}
\includegraphics[width=0.45\textwidth]{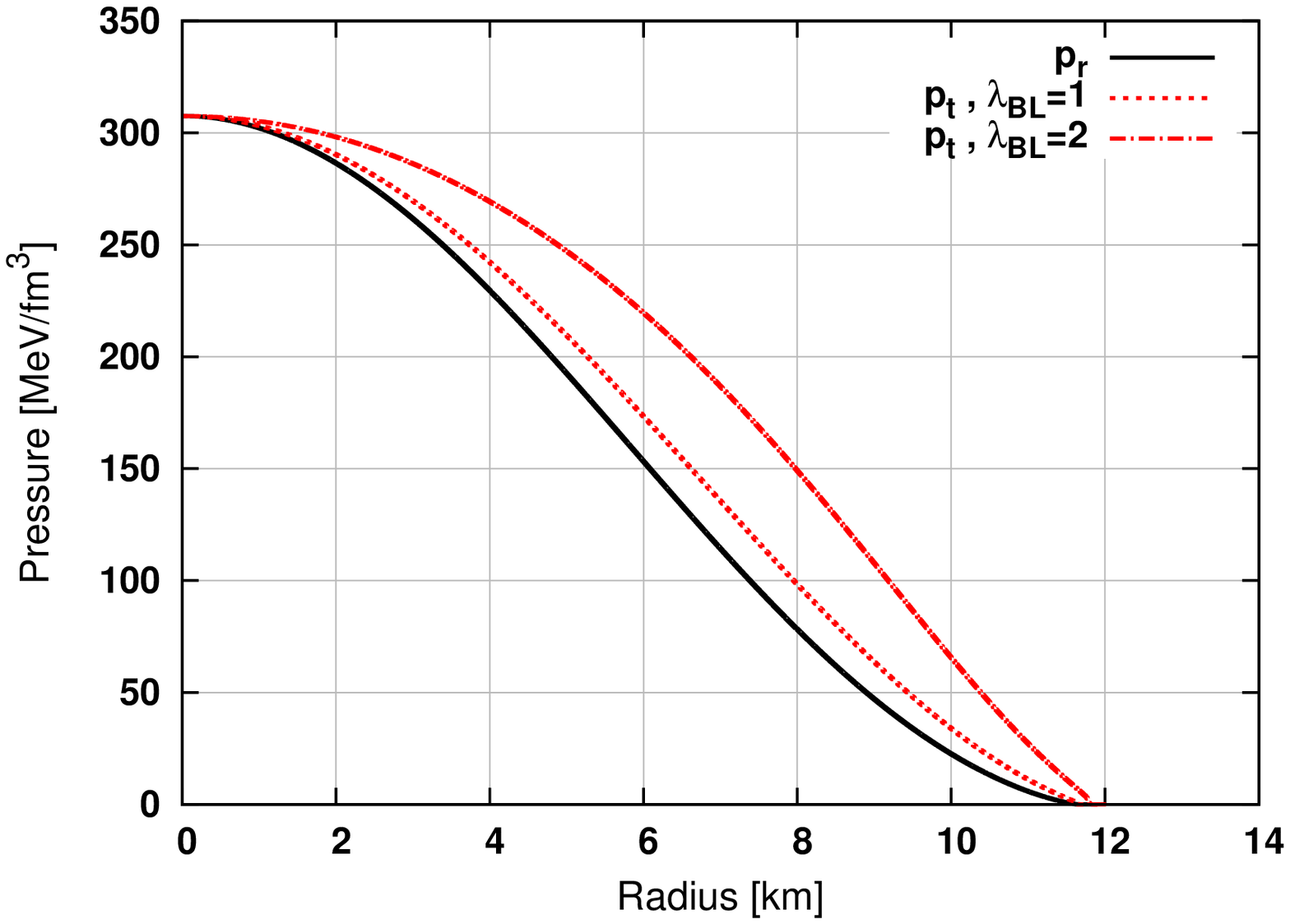}
\end{center}
\caption{Comparison of radial pressure and transverse pressure as functions of $r$ for $\lambda_{BL}=1$ and $\lambda_{BL}=2$ using DDH$\delta$ EOS for the central density $\rho= 5 \rho_0$. In both cases, the transverse pressure 
remains within a factor of a few of
the radial pressure and does not violate causality. As we study later, even smaller values of the transverse pressure can have observational consequences.}

\label{fig:pressure comparison}
\end{figure*}

In Fig.~\ref{fig:pressure comparison} both radial pressure and transverse pressure are plotted as functions of $r$ for the central baryon density $\rho =5 \rho_0$ (chosen arbitrarily) using DDH$\delta$ EOS for the positive values of $\lambda_{BL}=1$ and 2. In both cases, we see transverse pressure does not differ drastically from the radial pressure. Therefore, the amount of anisotropy that is allowed in this work is quite reasonable and also does not violate causality.

\section{COMPUTATION OF TIDAL DEFORMABILITY}
\label{sec3}

\begin{figure*}[ht]
\begin{center}
\begin{tabular}{cc}
\includegraphics[width=0.45\textwidth]{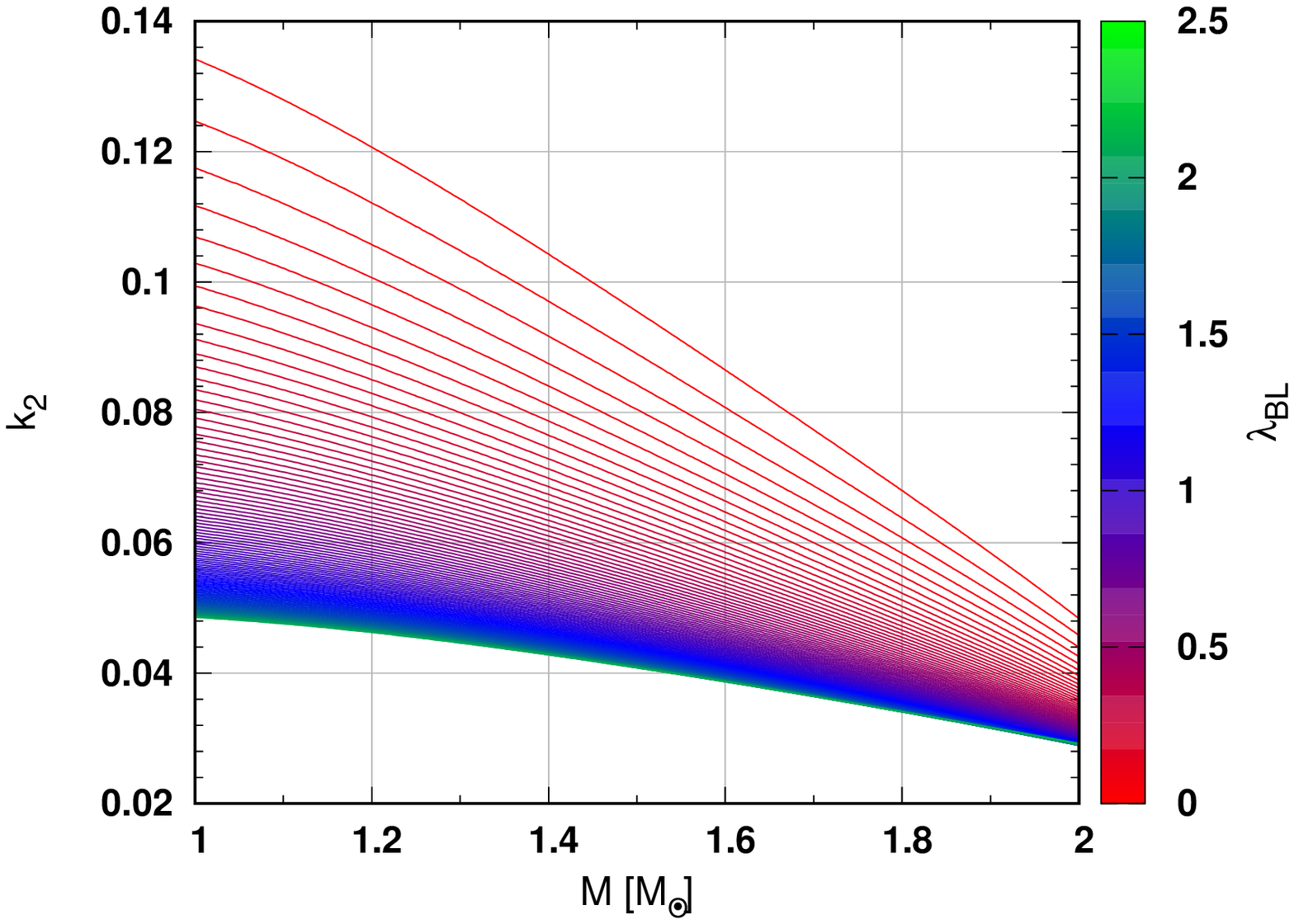}&
\includegraphics[width=0.45\textwidth]{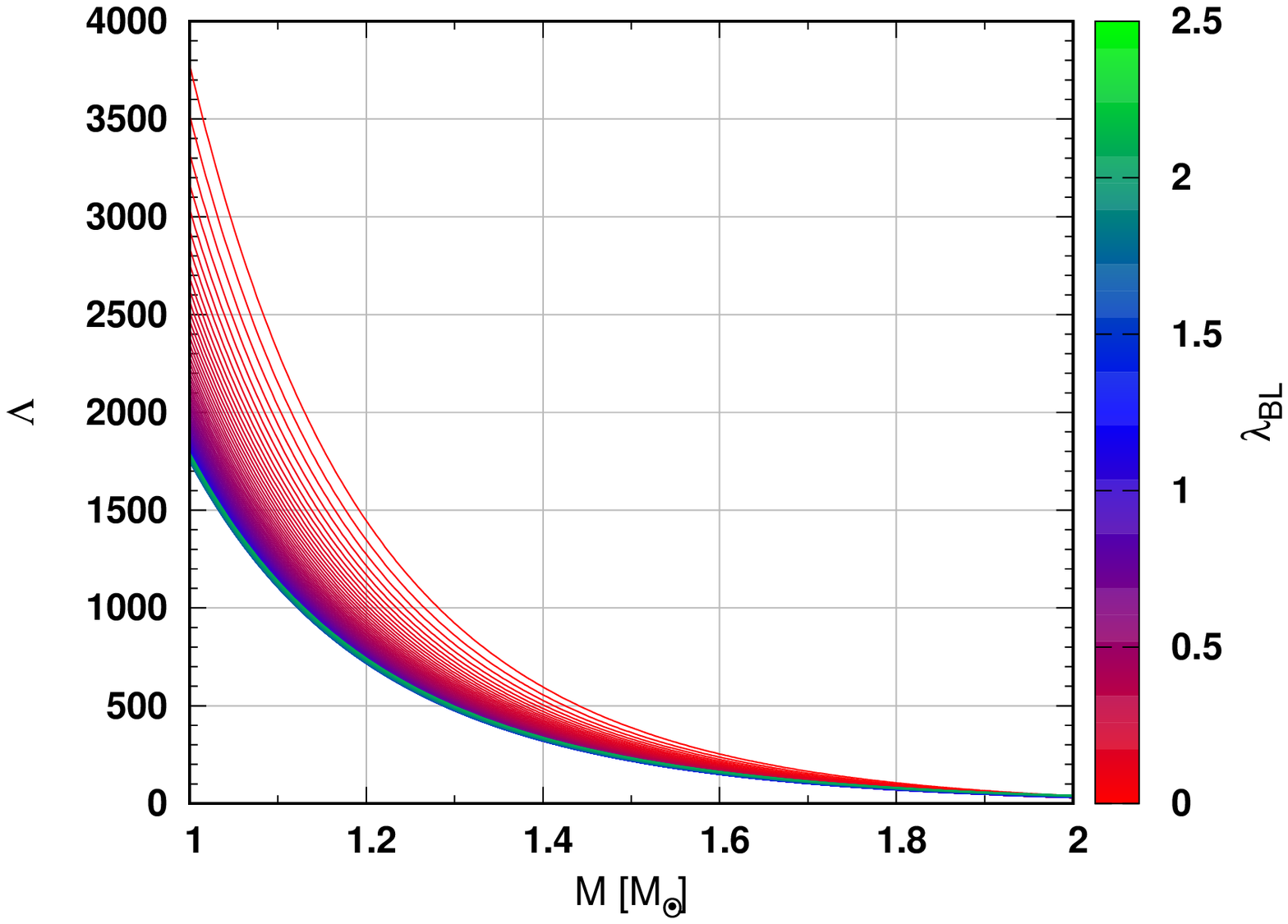}\\
\includegraphics[width=0.45\textwidth]{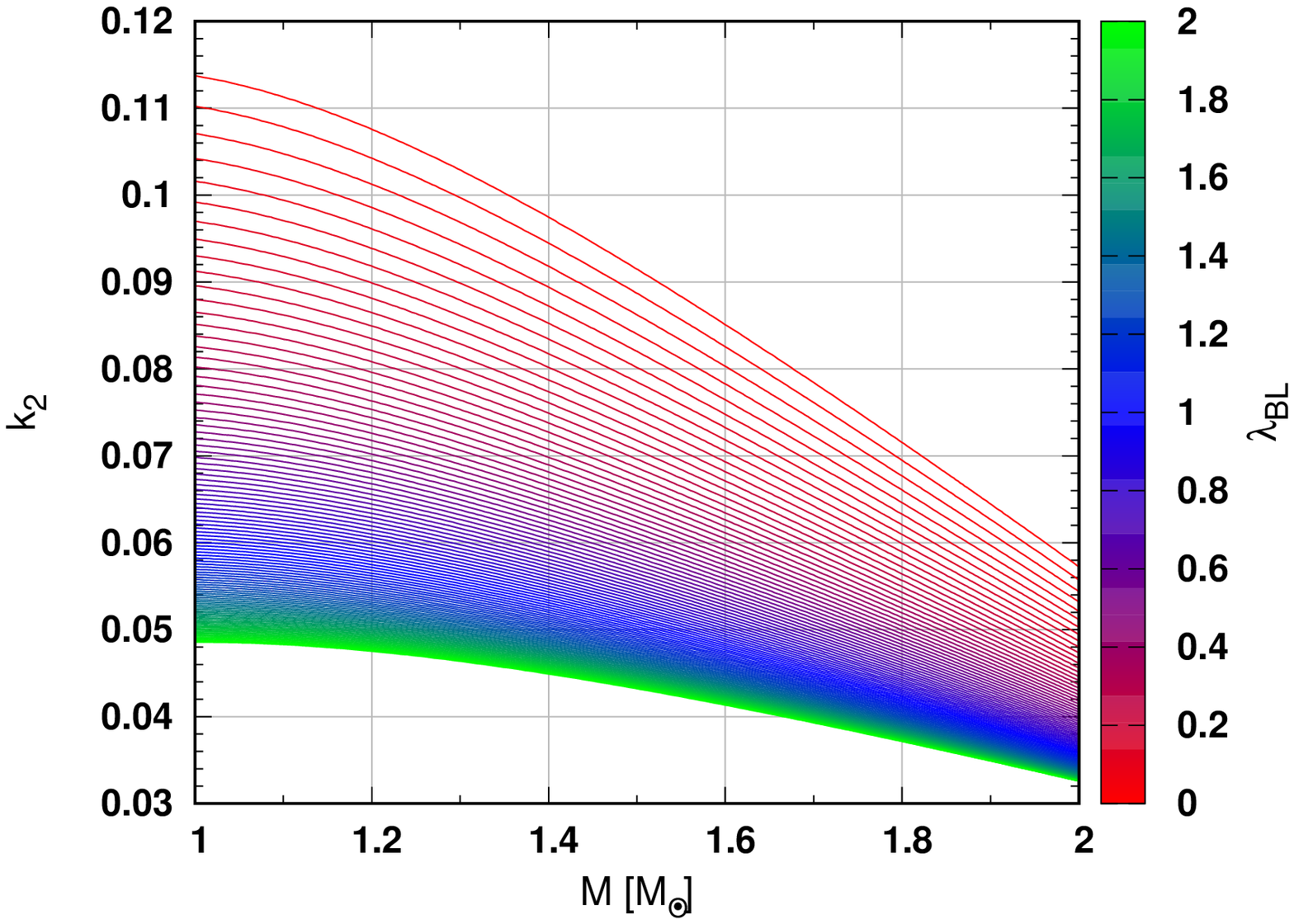}&
\includegraphics[width=0.45\textwidth]{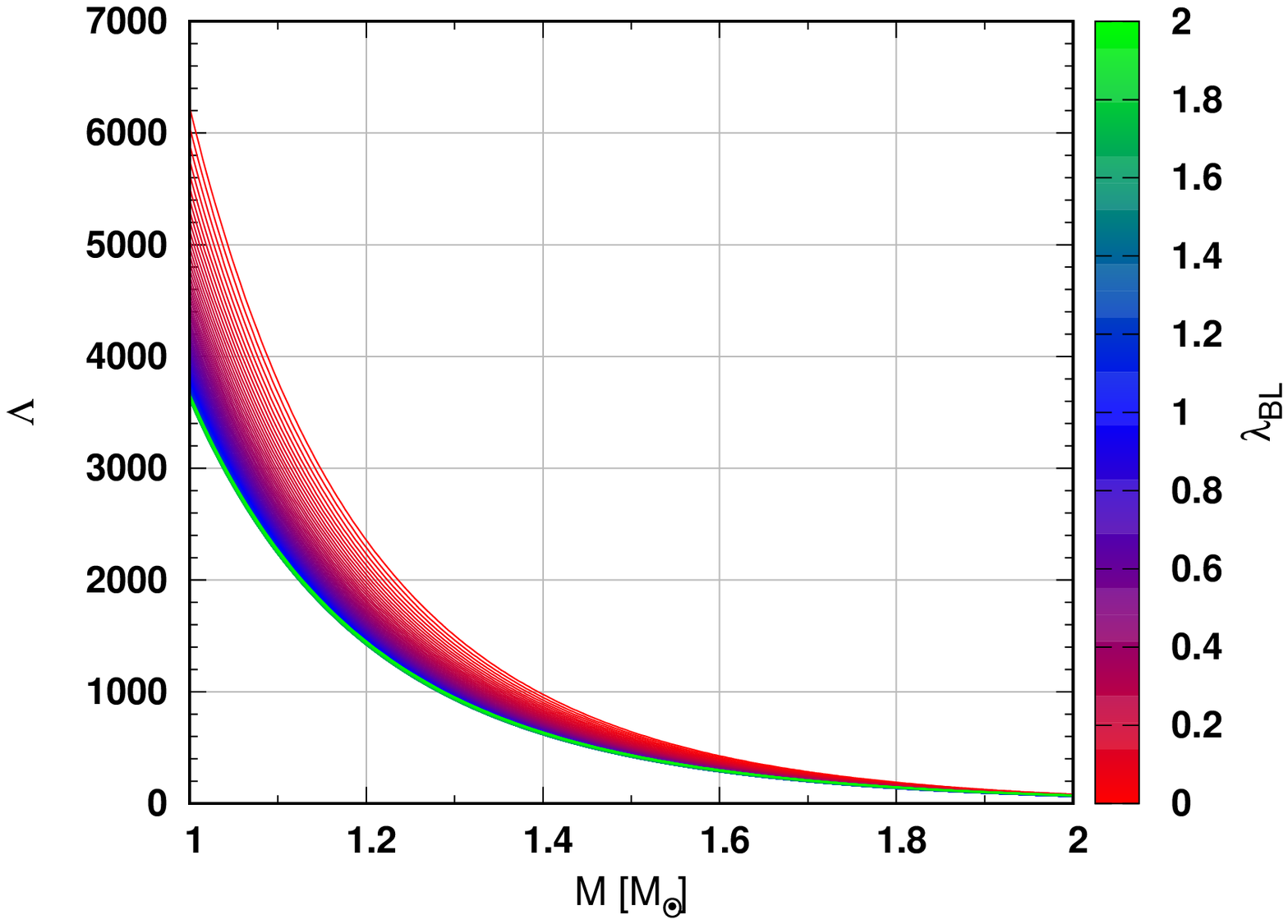}\\
\end{tabular}
\end{center}
\caption{Tidal Love number $k_2$ (left panel) and dimensionless tidal deformability $\Lambda$ (right panel) are plotted as functions of mass using EOSs DDH$\delta$ (top panel) and GM1 (bottom panel) for positive values of $\lambda_{BL}$.}
\label{fig:Love}
\end{figure*}

In the presence of an external tidal field $\epsilon_{ij}$ the equilibrium configuration of a neutron star gets tidally deformed. As a result the spherically symmetric star develops a quadrupole moment $Q_{ij}$. To linear order in $\epsilon_{ij}$, this induced response of the body is described as
\begin{equation}
Q_{ij}=-\lambda \epsilon_{ij}\,,
\end{equation}
where $\lambda$ is tidal deformability of the neutron star and is related to the dimensionless second Love number $k_2$ as $\lambda=\frac{2}{3}k_2 R^5$.
We also denote the mass of the star as $M$. To determine $k_2$, we study linear perturbation of the background metric following Thorne and Campolattaro~\cite{Thorne1967}:
\begin{equation}
g_{\alpha \beta}=g_{\alpha \beta}^{(0)} + h_{\alpha \beta}\,,
\end{equation}
where $h_{\alpha \beta}$ is the linearized perturbed metric. We expand components of metric and fluid perturbation variables in terms of spherical harmonics $Y_{lm}$ \cite{Regge wheeler 1957}. We restrict ourselves to the static $l=2,m=0$ even parity perturbations. With these restrictions the perturbed metric becomes
\begin{equation} 
   h_{\alpha \beta}={\rm diag}
\left[~ H_0(r) e^\nu, ~ H_2(r) e^\lambda, ~ r^2 K(r), ~
r^2 \sin^2\theta K(r)\right] Y_{2m}(\theta, \varphi)\,,
   \label{Metric_Perturbed}
\end{equation}
where $H_0$, $H_2$, and $K$ are all radial functions determined by the perturbed Einstein equations.

Expansion of the perturbed stress-energy tensor gives us the following relations: $\delta T_0^0=\delta \rho =\frac{d \rho}{dp} \delta p$, $\delta T_r^r =-\delta p$ and $\delta T_{\theta}^{\theta}=\delta T_{\varphi}^{\varphi}=-\delta p_t=-\frac{dp_t}{dp}\delta p$. We insert these fluid and metric perturbations in linearized Einstein equations $\delta G_{\alpha}^{\beta}=8 \pi \delta T_{\alpha}^{\beta}$. From $\delta G_{\theta}^{\theta}-\delta G_{\varphi}^{\varphi}=0$ and $\delta G_{\theta}^r=0$ we get $H_0=H_2\equiv H$ and $K^{'}=H \nu^{'}+H^{'}$, respectively. By subtracting the equation $\delta G_{\theta}^{\theta}+\delta G_{\varphi}^{\varphi}=-16 \pi \delta p_t$ from the tt-component of the perturbed Einstein equations we obtain the following differential equation for $H$:
	\begin{equation}\label{Anisotropic_Fluid_Perturbation}
	H^{''}+ H^{'} \bigg[\frac{2}{r} + e^{\lambda} \left(\frac{2m(r)}{r^2} + 4 \pi r (p - \rho)\right)\bigg] +  
	 H \left[4\pi e^{\lambda} \left(4 \rho + 8 p + \frac{\rho + p}{A {c_s}^2}(1+c_s^2)\right) -\frac{6 e^{\lambda}}{r^2} - {\nu^\prime}^2\right] = 0\,,
	\end{equation}
where $A\equiv \frac{dp_t}{dp}$, and $c_s^2\equiv \frac{dp}{d \rho}$ is the speed of sound squared; moreover, the prime denotes derivative with respect to $r$. If we put $A=1$ in Eq.~(\ref{Anisotropic_Fluid_Perturbation}) we recover the familiar master equation for the isotropic case~\cite{Hinderer2008}.

 Tidal Love number can be calculated by matching the internal solution with the external solution of the perturbed variable $H$ at the surface of the star \cite{Hinderer2008,Binnington2009,Damour2009}. Then the value of tidal Love number can be found in terms of $y$ and compactness parameter $C=\frac{M}{R}$:
\begin{widetext}
\begin{equation}
\label{expr_k2}
\begin{split}
k_2 &= \frac{8}{5}(1-2C)^2C^5\big[2C(y-1)-y+2\big]\bigg[2C(4(y+1)C^4 + (6y-4)C^3+(26-22y)C^2\\
&+3(5y-8)C-3y+6)-3(1-2C)^2(2C(y-1)-y+2)\log(\frac{1}{1-2C})\bigg]^{-1}\,,
\end{split}
\end{equation}
\end{widetext}
where $y$ depends on the value of $H$ and its derivative at the surface:
\begin{center}
$y=\left.\frac{rH^{'}}{H}\right\vert_{R}$\,.
\end{center}
In the left panel of Fig.~\ref{fig:Love}, the tidal Love number $k_2$ is plotted as a function of mass for positive $\lambda_{BL}$ using DDH$\delta$ (top panel) and GM1 (bottom panel). The isotropic case corresponds to $\lambda_{BL}=0$. We observe that as $\lambda_{BL}$ increases, the tidal Love number at a constant stellar mass decreases for both EOSs. In the right panel of Fig.~\ref{fig:Love} the dimensionless tidal deformability $\Lambda \equiv \lambda /M^5$ is plotted as a function of the star's mass between 1$M_{\odot}$ to 2$M_{\odot}$. We observe that positive anisotropy reduces the value of $\Lambda$, for a given mass. 

\section{Implications of GW170817 on EOS with anisotropic pressure}
\label{sec4}

\begin{figure*}[ht]
\begin{center}
\begin{tabular}{cc}
\includegraphics[width=0.45\textwidth]{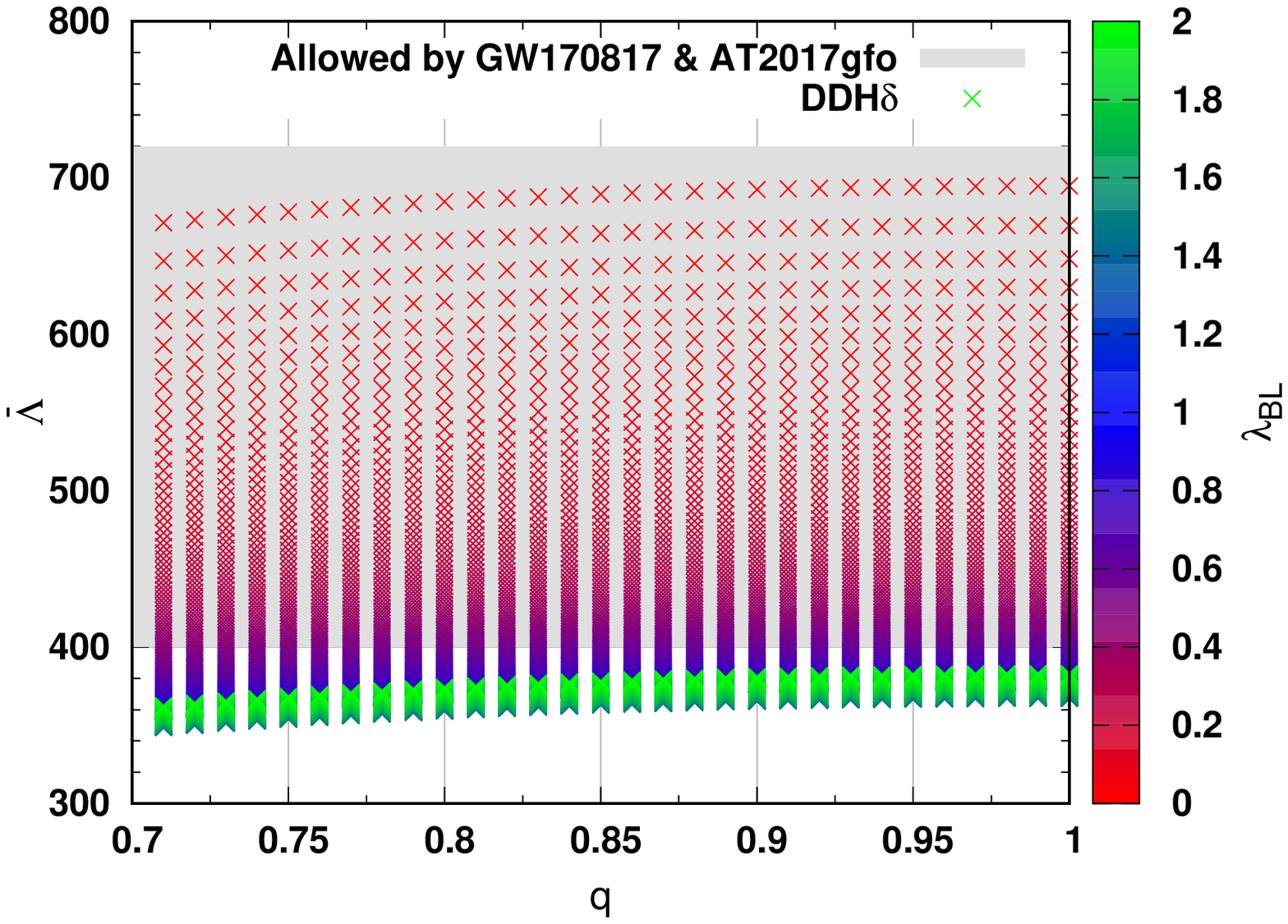}&
\includegraphics[width=0.45\textwidth]{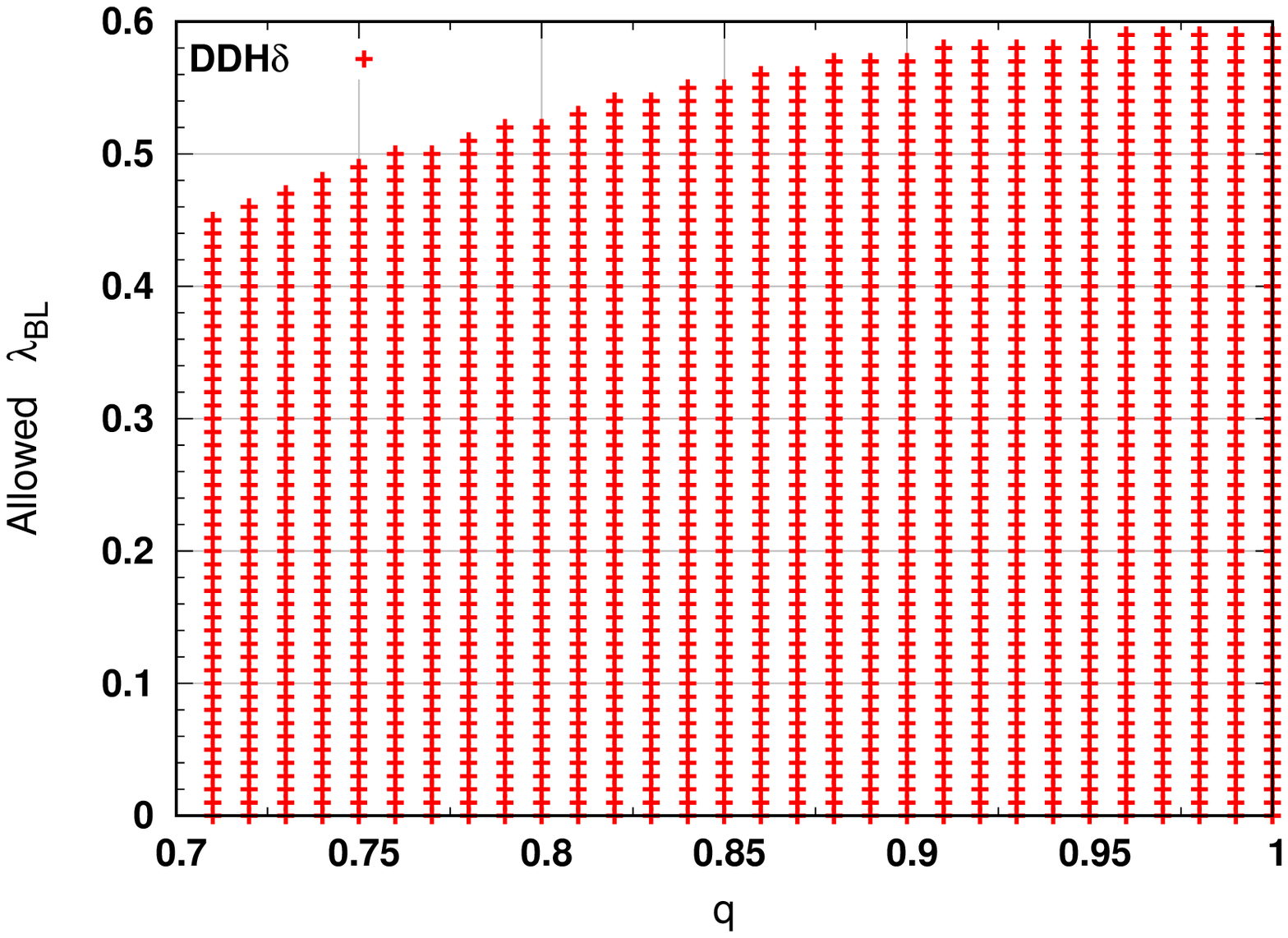}\\
\includegraphics[width=0.45\textwidth]{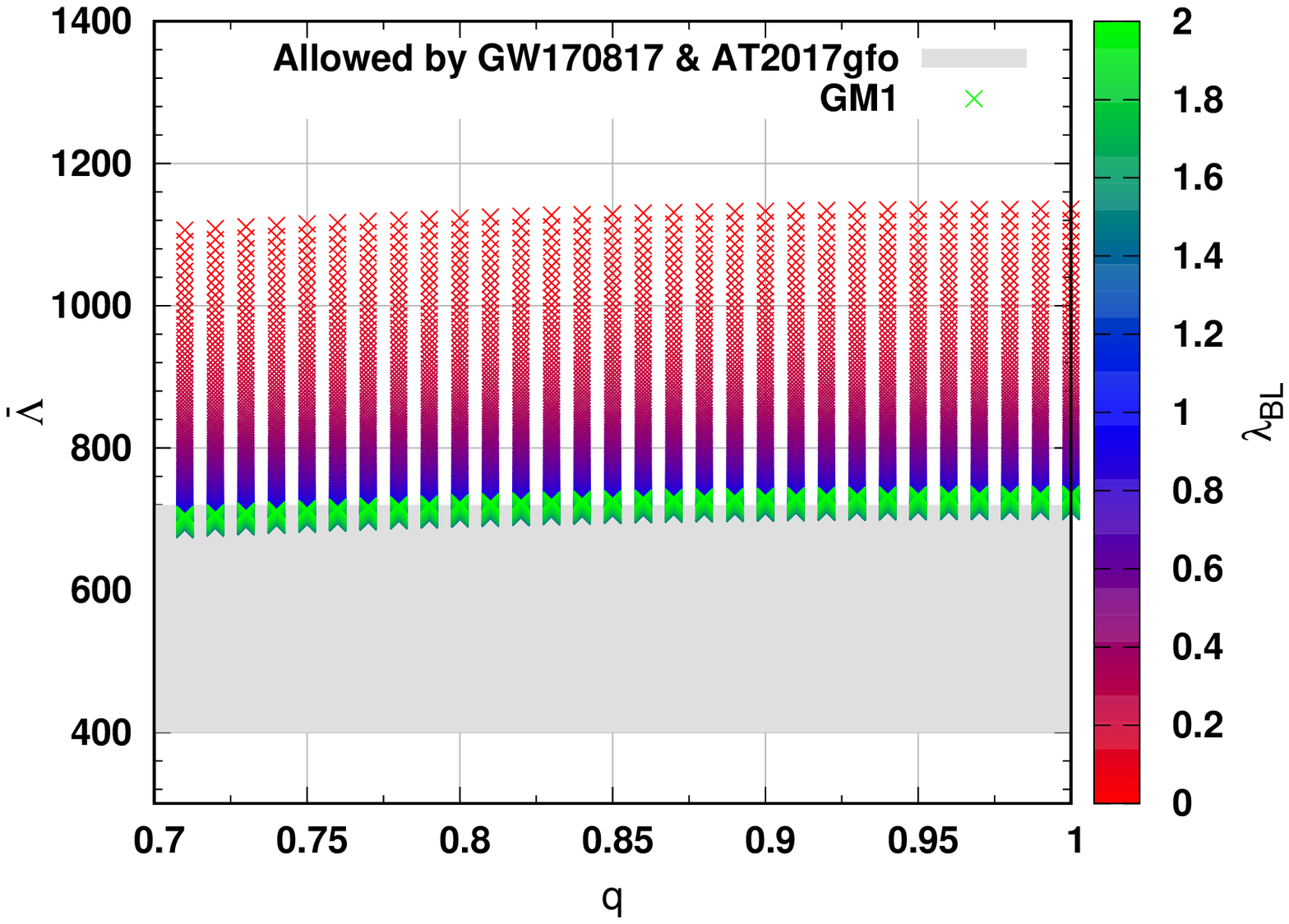}&
\includegraphics[width=0.45\textwidth]{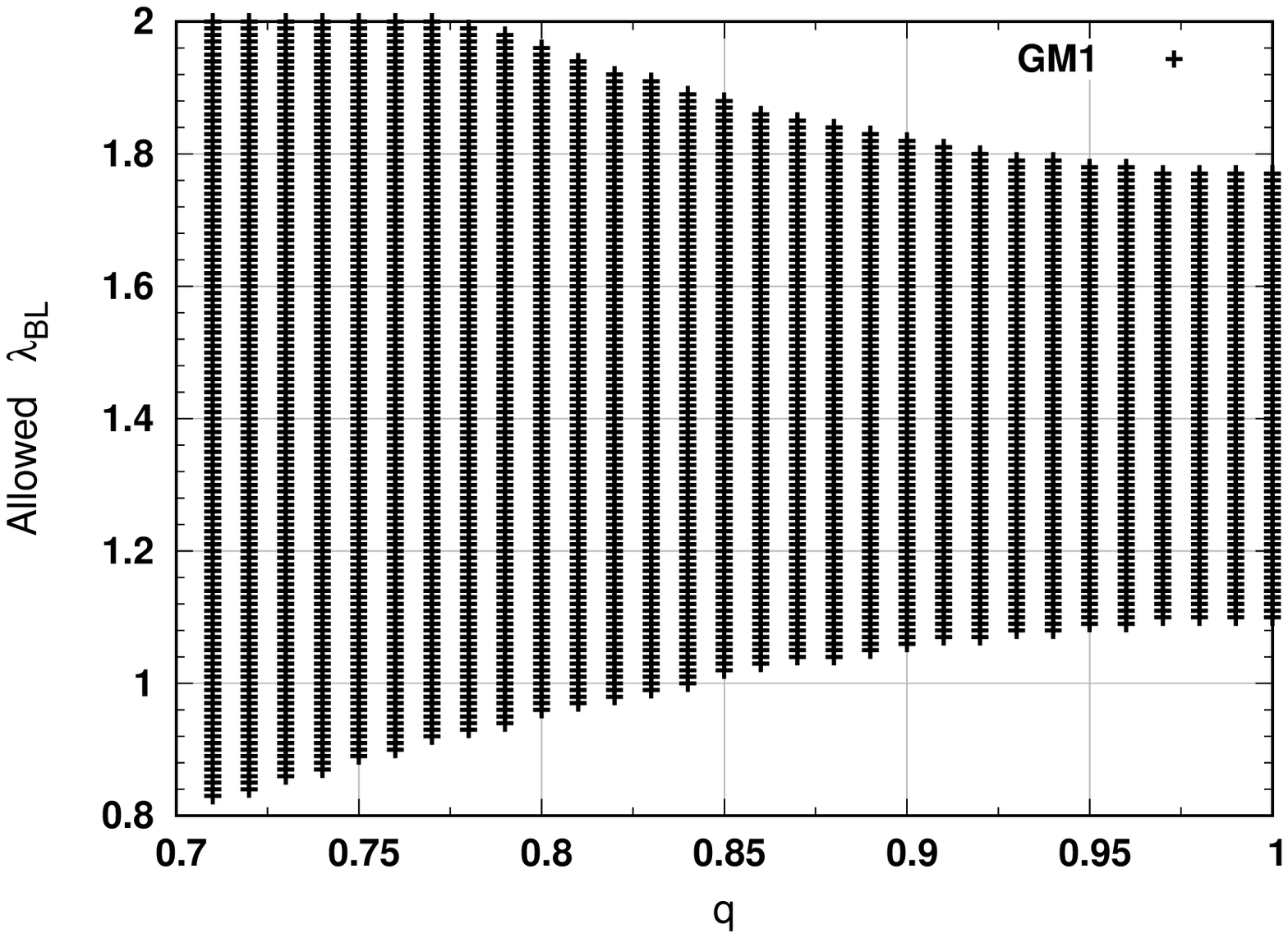}\\
\end{tabular}
\end{center}
\caption{$\bar{\Lambda}$ (left panel) and allowed values of $\lambda_{BL}$ are plotted as functions of the mass ratio $q$ for DDH$\delta$ (upper panel) and GM1 (lower panel) EOSs. The shaded regions in the left panel are allowed by GW-EM observations of GW170817 and AT2017gfo.}
\label{fig:Allowed_BL}
\end{figure*}

At leading order $\Lambda_{1,2}$ appear in the gravitational wave phase through the effective tidal deformability
\begin{equation}    \bar{\Lambda}=\frac{16}{13}\frac{(m_1+12m_2)m_1^4\Lambda_1+(m_2+12m_1)m_2^4\Lambda_2}{(m_1+m_2)^5} \,,
\end{equation}
where $\Lambda_1$ and $\Lambda_2$ are the tidal deformabilities of the heavier and lighter stars, respectively. The recent detection of GWs from the binary neutron star merger event GW170817 has 
constrained $\bar{\Lambda}$ to be $\leq 720$~\cite{Abbott:2018exr,properties} at 90\% confidence level for low spin (dimensionless spin magnitude $\leq 0.05$) prior. The corresponding chirp mass, $M_c=\frac{(m_1m_2)^{3/5}}{(m_1+m_2)^{1/5}}$, was measured to be 
$1.188^{+.004}_{-.002} M_{\odot}$ and the mass ratio, $q=\frac{m_1}{m_2}$,
was constrained between 0.7-1 for low spin prior. Also the EM counterpart of GW170817, named AT2017gfo, provides an additional constraint of $\bar{\Lambda} \geq 400$~\cite{Radice2018}. Indeed, constraints from chiral effective field theory and perturbative quantum chromodynamics suggest for a lower bound on $\Lambda$ that is as low as 120 for 1.4$M_\odot$~\cite{Annala:2017llu}. As we show below, a lower value of $\Lambda_{1,2}$ and, therefore, $\bar{\Lambda}$, can allow for a larger range of $\lambda_{BL}$ to be admissible by GW170817 observations. Here, we choose to be conservative and take $\bar{\Lambda} \geq 400$. Combining these GW and EM constraints gives the allowed range of $\bar{\Lambda}$ to be $400 \leq \bar{\Lambda} \leq 720$.

Many of the relativistic equations of state struggle to satisfy the upper bound, $\bar{\Lambda} \leq 720$ (see, e.g., Ref.~\cite{Ranada2018}), assuming $\lambda_{BL}=0$. That situation changes if anisotropy in pressure is present. In the left panel of Fig.~\ref{fig:Allowed_BL}, we have plotted $\bar{\Lambda}$ as a function of $q$ for positive $\lambda_{BL}$ using both DDH$\delta$ (upper panel) and GM1 (lower panel) EOSs. The allowed ranges of $\bar{\Lambda}$ are shaded in gray. Figure~\ref{fig:Allowed_BL} shows that between GM1 and DDH$\delta$, only the latter satisfies the constraint on $\bar{\Lambda}$ set in Ref.~\cite{Abbott:2018exr,properties} when $\lambda_{BL}=0$. Indeed, $\bar{\Lambda} > 1000$ if these were GM1 stars with $\lambda_{BL}=0$.
However,  when $\lambda_{BL} \neq 0$ the value of $\bar{\Lambda}$ falls by a large amount for both EOSs, so much so that it can lie within the GW-EM bounds for a certain amount of pressure anisotropy.
Furthermore, the lower bound on $\bar{\Lambda}$ helps limit the value of $\lambda_{BL}$ from above.
In the right panel of Fig.~\ref{fig:Allowed_BL} the allowed ranges of $\lambda_{BL}$ are plotted against $q$ for DDH$\delta$ (upper panel) and GM1 (lower panel) EOSs. 
We find that presence of anisotropy in pressure can reduce the value of $\Lambda$ by a significant amount. Thus, certain EOSs that were ruled out by GW170817 observations for $\lambda_{BL}=0$ become viable if the stars support an anisotropic component in the pressure.

\subsection{Using universality relations for further constraining pressure anisotropy}
\label{universal}

\begin{figure*}[ht]
\begin{center}
\begin{tabular}{cc}
\includegraphics[width=0.5\textwidth]{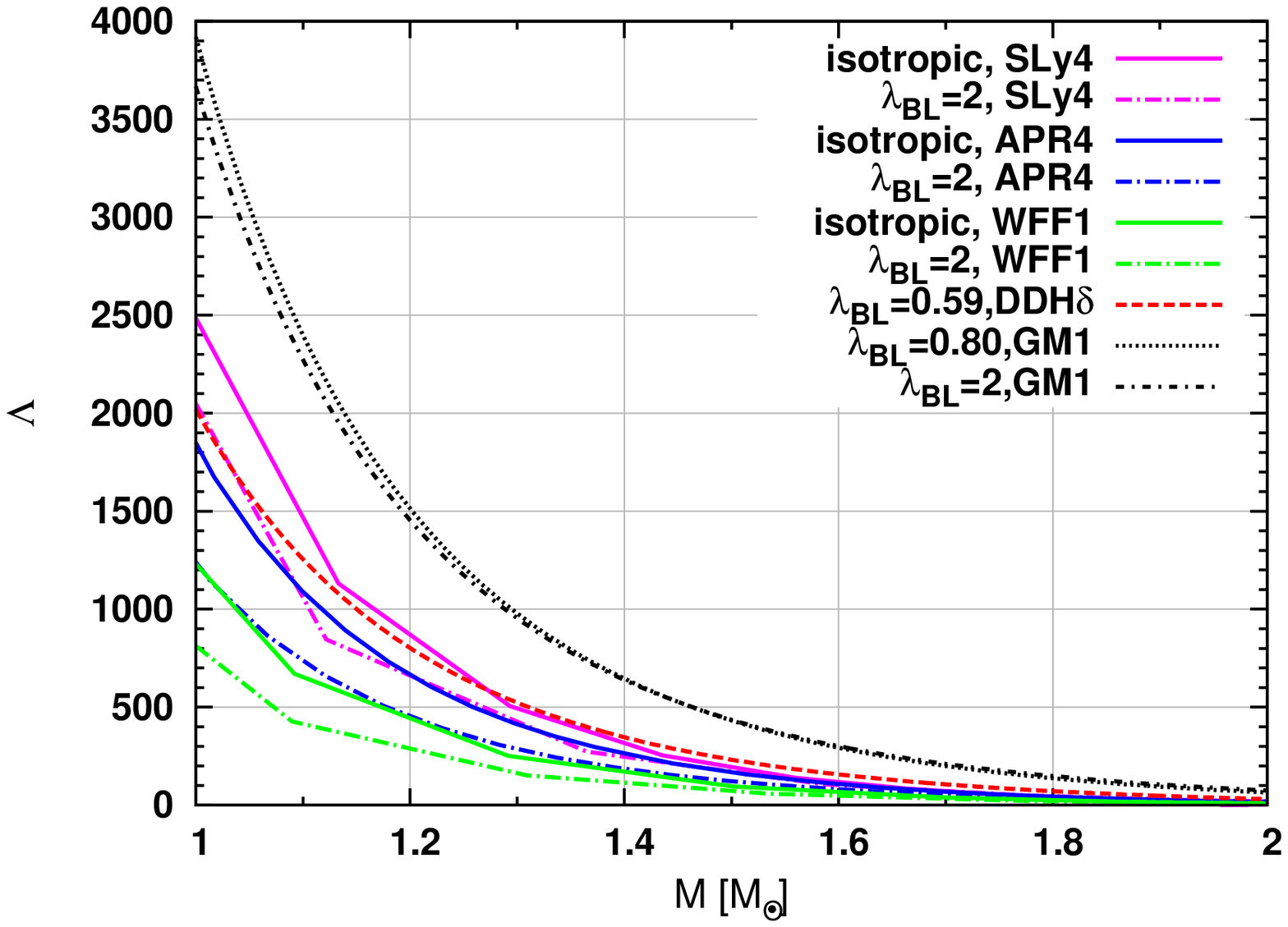}&
\includegraphics[width=0.5\textwidth]{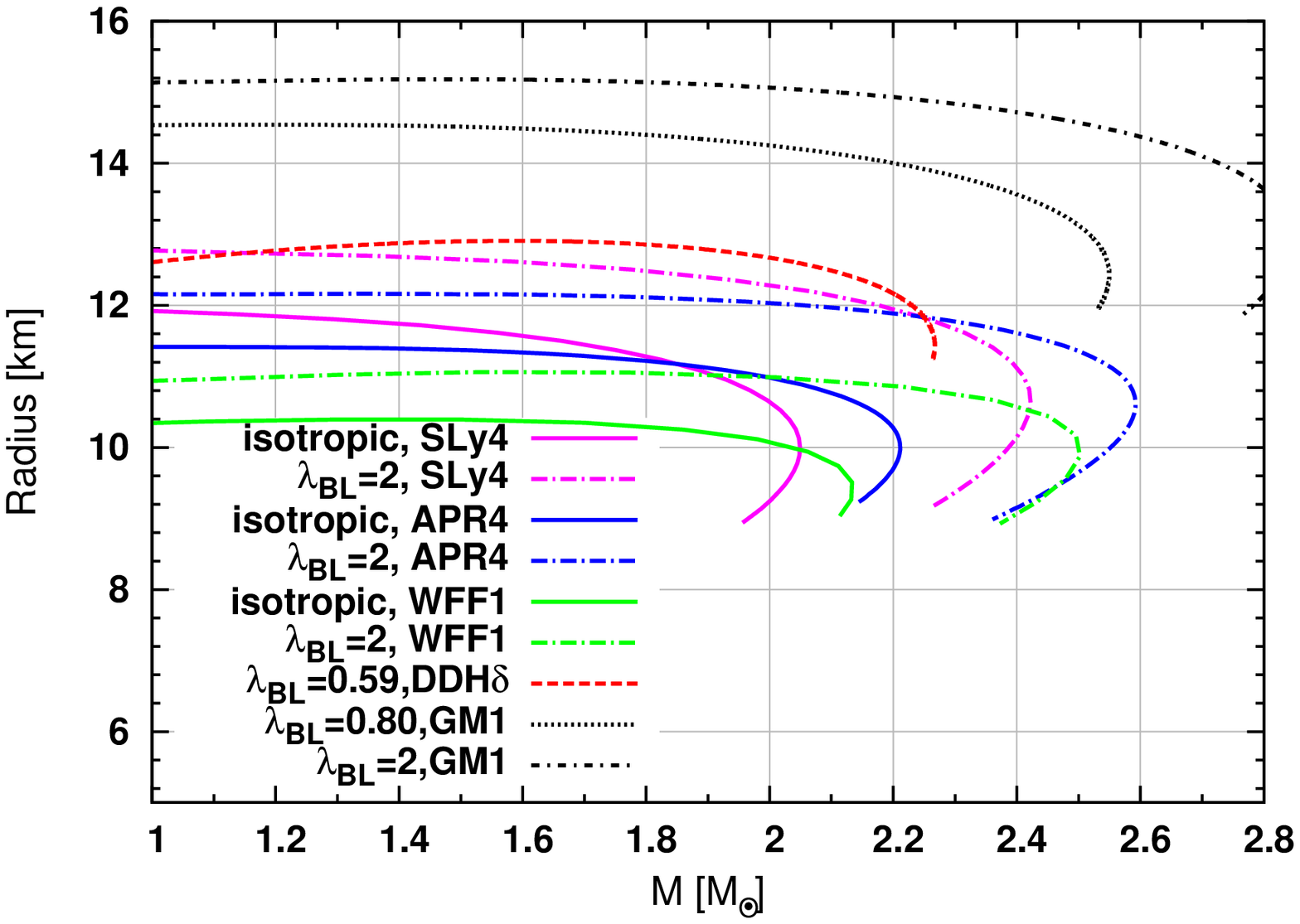}\\
\end{tabular}
\end{center}
\caption{Dimensionless tidal deformability $\Lambda$ (left panel) and radius (right panel) is plotted as a function of mass: The solid lines correspond to isotropic neutron stars (black, blue and green are respectively for Sly4, Apr4 and Wff1 EOSs) and the red dotted line corresponds to  DDH$\delta$ stars but with EOS anisotropic pressure $\lambda_{BL}=.26$.
}
\label{fig:degenneracy break}
\end{figure*}

As shown in Fig.~\ref{fig:Love} in the mass range of interest $[1,2]M_\odot$,
in the presence of positive pressure anisotropy
the tidal Love number decreases for any fixed stellar mass. Thus, the $\Lambda$ distribution of stars with a soft EOS, such as SLy4 with no pressure anisotropy, can be difficult to distinguish from that of stars with a stiff EOS, such as DDH$\delta$, but non-zero pressure anisotropy, say, $\lambda_{BL}=.59$; see the left panel of Fig.~\ref{fig:degenneracy break}.
In this sense, positive anisotropy has an effect that is similar to making a star softer, for a given mass.
This poses the problem of how one might distinguish these two types of stars. We argue here that it is possible to make the correct identification in some cases by measuring the stellar radius. This is because a non-zero $\lambda_{BL}$ tends to make the star larger, for any fixed stellar mass (see the right panel of Fig.~\ref{fig:degenneracy break}). Note there that DDH$\delta$ with $\lambda_{BL}=.59$ has a larger radius for most of the mass range than SLy4 with any $\lambda_{BL} \in [0,2]$.

\begin{figure*}[ht]
\begin{center}
\begin{tabular}{cc}
\includegraphics[width=0.45\textwidth]{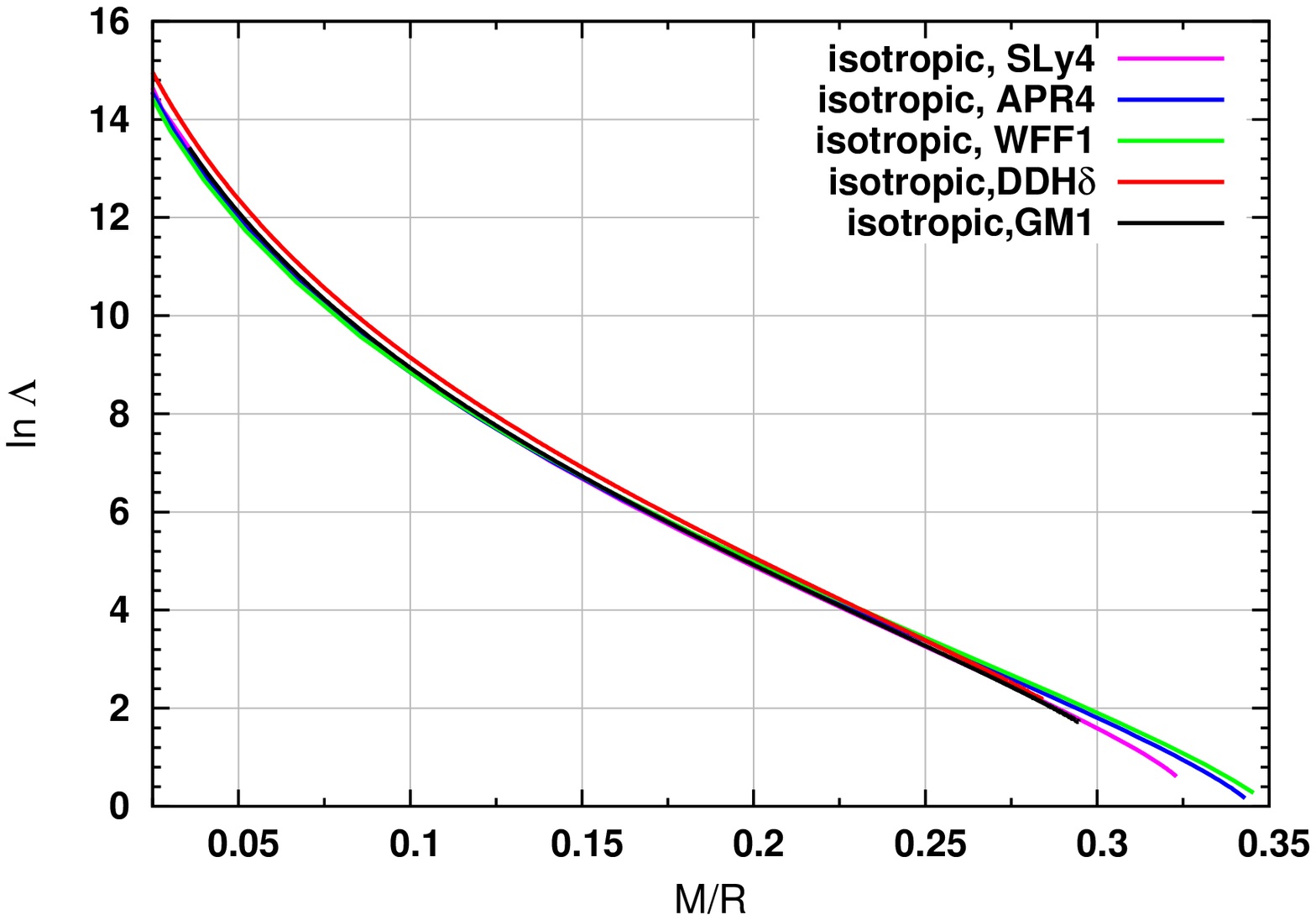}&
\includegraphics[width=0.45\textwidth]{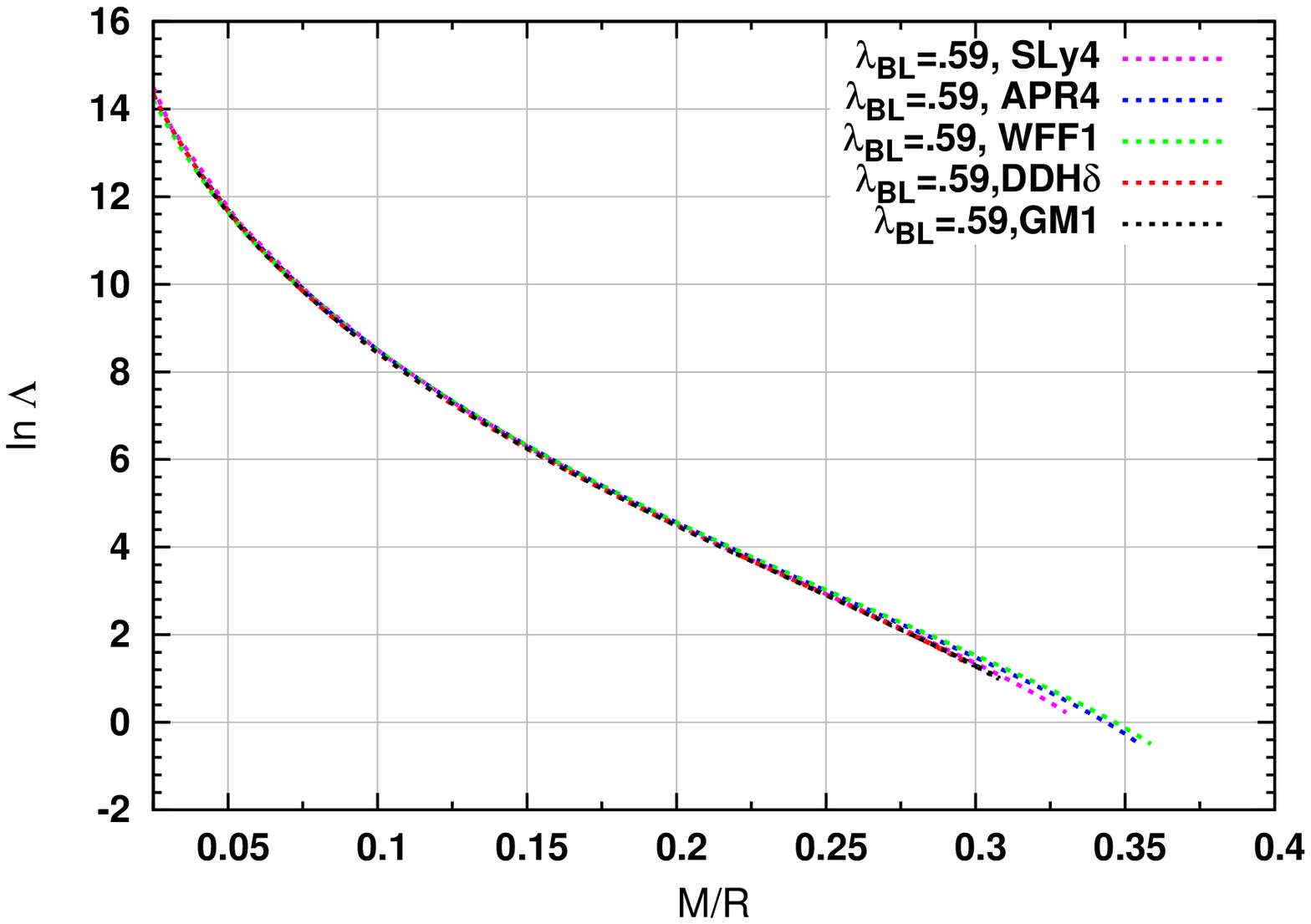}\\
\includegraphics[width=0.45\textwidth]{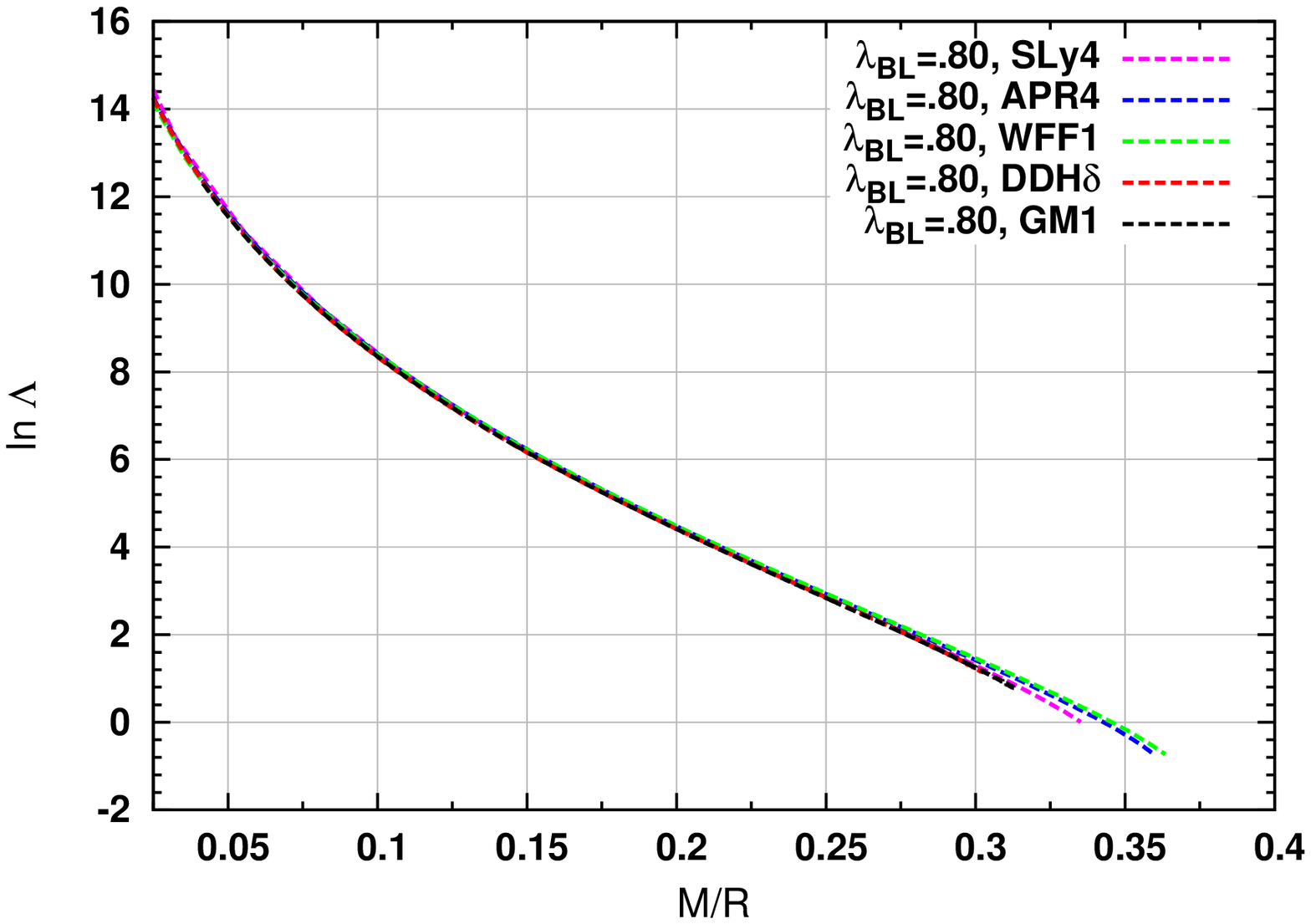}&
\includegraphics[width=0.45\textwidth]{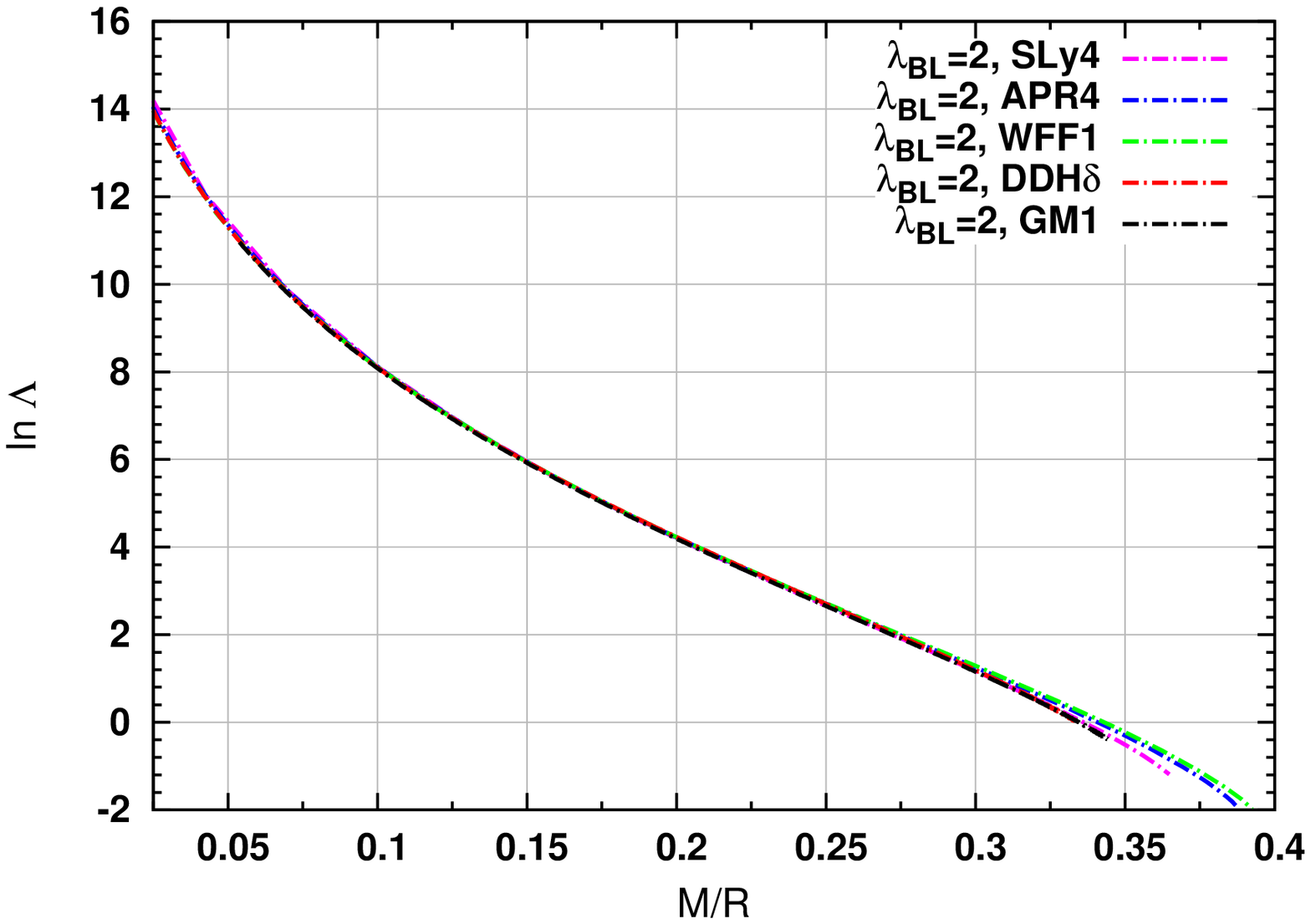}\\
\end{tabular}
\end{center}
\caption{$C-\Lambda$ relationship for different cases : different EOSs with $\lambda_{BL}=0$ (top left), different $\lambda_{BL}$ for fixed DDH$\delta$ EOS (top right), different EOSs with $\lambda_{BL}=1$ (bottom left), different EOSs with $\lambda_{BL}=2$ (bottom right).}
\label{fig:Universal relation}
\end{figure*}

To measure the stellar radius, we adopt the same trick that was resorted to in Ref.~\cite{Abbott:2018exr}, namely, to use universality relations between $\Lambda$ and stellar compactness. 
Universality of the $C-\Lambda$ relationship was first pointed out by Maselli et al.~\cite{Maselli2013}.  Here, we inspect whether this universality also holds in the presence of pressure anisotropy, Eq.~(\ref{Anisotropy_eos}). In the top left panel of Fig.~\ref{fig:Universal relation}, $C$ vs $\ln \Lambda$ is plotted for five different isotropic EOSs. We find that $C$ is well fitted by the following relation,
\begin{equation}
    C=0.356883 -0.0363734 \ln \Lambda + 0.000899844 (\ln \Lambda)^2 \,.
\end{equation}
GW170817 has constrained the value of $\Lambda$ for a 1.4 $M_{\odot}$ star to be $190^{+390}_{-120}$~\cite{Abbott:2018exr}. If we use the above mentioned $C-\Lambda$ relationship and this constraint on $\Lambda$, then the radius of the two stars in GW170817 is measured to be $10.8^{+2.0}_{-1.4}$km, with the upper limit consistent with Ref.~\cite{Abbott:2018exr}, and the lower limit larger by 0.3km at most.

In the top right panel of Fig.~\ref{fig:Universal relation}, $C$ vs $\ln \Lambda$ is plotted for different EOSs but with fixed $\lambda_{BL}=0.59$. We observe universality relation holds, in fact somewhat more tightly, for such pressure anisotropy.
The fitted $C-\Lambda$ relation for this anisotropic configuration is:
\begin{equation}
\label{univ-compactness}
    C=0.349945-0.0382978 \ln \Lambda + 0.00106643 (\ln \Lambda)^2 \,,
\end{equation} 
and the radius is constrained to $11.6^{+2.2}_{-1.6}$ km for the same $\Lambda$ measurement of $190^{+390}_{-120}$.
With $\lambda_{BL}=0.80$ (bottom left panel),
\begin{equation}
\label{univ-compactness-0p80}
    C=0.347709-0.0384087 \ln \Lambda + 0.00108876 (\ln \Lambda)^2 \,,
\end{equation}
and the radius will be constrained to $11.7^{+2.3}_{-1.6}$ km for the aforementioned $\Lambda$.
With $\lambda_{BL}=2$ (bottom right panel),
\begin{equation}
    C=0.337864-0.0367376 \ln \Lambda + 0.000985149 (\ln \Lambda)^2 \,,
\end{equation}
and the radius is found to be $12.0^{+2.3}_{-1.6}$ km.

We illustrate in Table~\ref{tab1} how the above radius measurements can be used to rule in or out various EOSs with non-zero $\lambda_{BL}$.
For example, the universality relation Eq.~(\ref{univ-compactness}) implies that for DDH$\delta$ neutron stars with the same masses as GW170817, and $\lambda_{BL}=.59$, the radius must obey  $11.6^{+2.2}_{-1.6}$ km (90\% CL), which allows for the maximum radius of such stars to be 13.8 km. However, Fig.~\ref{fig:degenneracy break} shows that the minimum radius for such a star in the 1 - 2$M_\odot$ mass range is $R_{\rm min}= 12.9$ km, which is less than 13.8km. This is why we infer that DDH$\delta$ remains viable following the observation of GW170817 provided $\lambda_{BL} \geq 0.59 $. Note that DDH$\delta$ with $\lambda_{BL} \geq 0.59 $ is allowed by the GW-EM constraint on $\bar{\Lambda}$, as already observed in Fig.~\ref{fig:Allowed_BL}.
On the other hand, the universality relation Eq.~(\ref{univ-compactness-0p80}) implies that for GM1 neutron stars with the same masses as GW170817, and $\lambda_{BL}=.80$, the radius must obey  $11.7^{+2.3}_{-1.6}$km (90\% CL), which allows for the maximum radius of such stars to be 14.0km. However, Fig.~\ref{fig:degenneracy break} shows that the minimum radius for such a star in the 1 - 2$M_\odot$ mass range is $R_{\rm min}= 14.2$km, which is larger than 14.0km. This is why we infer that GM1, with $\lambda_{BL} \geq 0.80 $, is ruled out following the observation of GW170817.

In the analysis of the GW170817 signal in Ref.~\cite{Abbott:2018exr},  LIGO and Virgo used another factor, arising from pulsar mass observations, namely that any viable EOS must support NS with a maximum mass that is at least 1.97 $M_{\odot}$. This requirement gives an improved measurement of radius. We leave the study of the corresponding impact in $\lambda_{BL}$ constraints to a future study.

\begin{table}[ht]
\caption{Use of radius constraints to discern the presence or absence of pressure anisotropy. For example, the universality relation Eq.~(\ref{univ-compactness}) implies that for DDH$\delta$  neutron stars with the same masses as GW170817, and $\lambda_{BL}=.59  $, the radius must obey  $11.6^{+2.2}_{-1.6}$km (90\% CL), which allows for the maximum radius of such stars to be 13.8km. However, Fig.~\ref{fig:degenneracy break} shows that the minimum radius for such a star in the 1 - 2$M_\odot$ mass range is $R_{\rm min}= 12.9$km, which is less than 13.8km. This is why we infer that DDH$\delta$ remains viable following the observation of GW170817 provided $\lambda_{BL} \geq 0.59 $.}
\begin{tabular}{cccccc} 
\hline
 EOS& Nature & Constraint on radius & $R_{\rm min}$ (km) & comment\\
\hline\hline 
 DDH$\delta$ &$\lambda_{BL}=.59  $ & $11.6^{+2.2}_{-1.6}$ & 12.7 & Survive \\
 \hline
GM1 & $\lambda_{BL}=.80$ & $11.7^{+2.3}_{-1.6}$ &14.2 & Ruled out \\
GM1 & $\lambda_{BL}=2$ & $12.0^{+2.3}_{-1.6}$ & 15.1 & Ruled out \\
\hline 
\end{tabular}
\label{tab1}
\end{table}

We also observe that more observations of neutron star mergers, as anticipated, will help constrain $\lambda_{BL}$ more tightly. This will help in narrowing the statistical errors, thereby allowing smaller systematic effects arising from $\lambda_{BL}$ to stand out.
 Indeed, when the statistical error of any GW observable gets so precise (e.g., with larger number of observations) that it is smaller than the systematic shift induced by non-zero $\lambda_{\rm BL}$, then it becomes meaningful to use it to constrain the presence of anisotropic pressure in these stars.

\section{CONCLUSION}
 
In this paper, we have calculated tidal Love number and deformability of neutron stars in the presence of anisotropic pressure. As a first step, we use two RMF EOSs to describe radial pressure and a functional form of anisotropic pressure as proposed by Bowers and Liang \cite{Bowers and Liang}. We obtain the equilibrium solutions numerically by integrating modified TOV Eqs. (\ref{tov1:eps}) and (\ref{tov2:eps}),
and find that they can differ significantly from the isotropic ones: We observe that for any fixed central density, the compactness of the star increases for positive anisotropy ($\lambda_{BL}>0$) and decreases for negative anisotropy ($\lambda_{BL}<0$). In a further investigation, when we plot transverse pressure as a function of radius for the chosen negative values of $\lambda_{BL}$ we notice that the $p_t\geq 0$ requirement is not met everywhere inside the star. Therefore, we discard those anisotropic EOSs with negative values of $\lambda_{BL}$.    

The tidal Love numbers and deformabilities are obtained by integrating the single equation (\ref{Anisotropic_Fluid_Perturbation}) for the perturbed metric variable $H$, along with the modified TOV Eqs. (\ref{tov1:eps}) and (\ref{tov2:eps}), and for the boundary condition $p(r=R)=0$. It turns out that both tidal Love numbers and deformabilities can reduce by a significant amount in the presence of pressure anisotropy. This leads to an interesting possibility. Earlier, a subset of those EOSs that failed to satisfy the bound of tidal deformability set by GW170817, can now become viable if anisotropy in pressure is present beyond a certain threshold. We demonstrate this by analyzing the cases of two RMF EOSs, DDH$\delta$ and GM1, for various values of  $\lambda_{BL}$. 
Finally we propose how future observations may be able to discern the presence or absence of anisotropic pressure.

\section*{ACKNOWLEDGMENT}

We thank Prasanta Char for providing the non-relativistic EOSs studied here.
We also would like to acknowledge Kabir Chakravarti, Sumanta Chakraborty and Sayak Datta for useful discussions. We thank Rahul Kashyap for reading the manuscript carefully.
This work was done in part with support from the Navajbai Ratan Tata Trust. B. Biswas acknowledges University grant commission of India, for the financial support as a senior research fellow. This document has been assigned the preprint number LIGO-DCC-P1800396.

\section*{APPENDIX: COMMENT ON ANISOTROPIC ULTRACOMPACT OBJECTS}
\label{sec:ECO}
 \begin{figure*}[ht]
\begin{center}
\begin{tabular}{cc}
\includegraphics[width=0.45\textwidth]{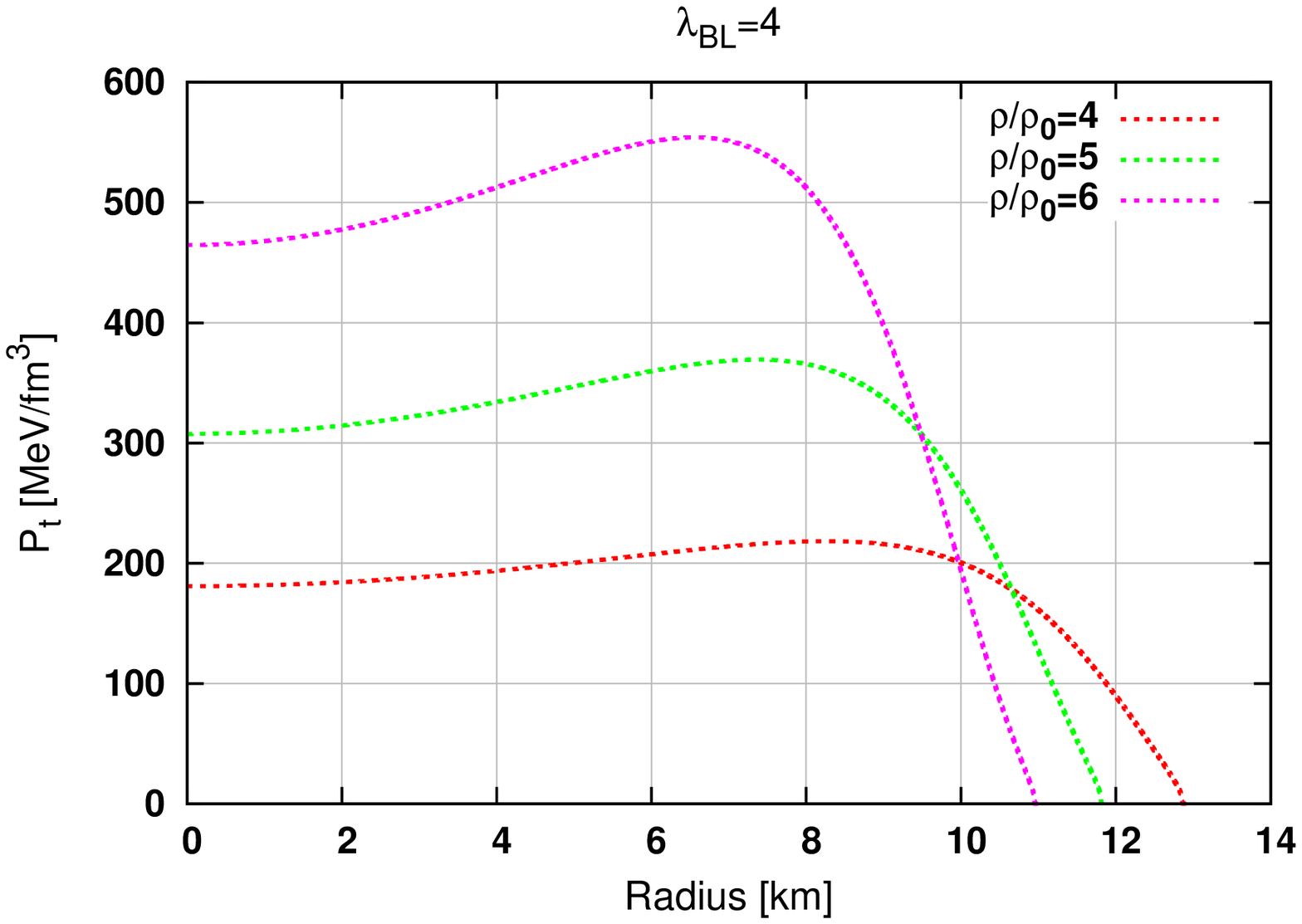}&
\includegraphics[width=0.45\textwidth]{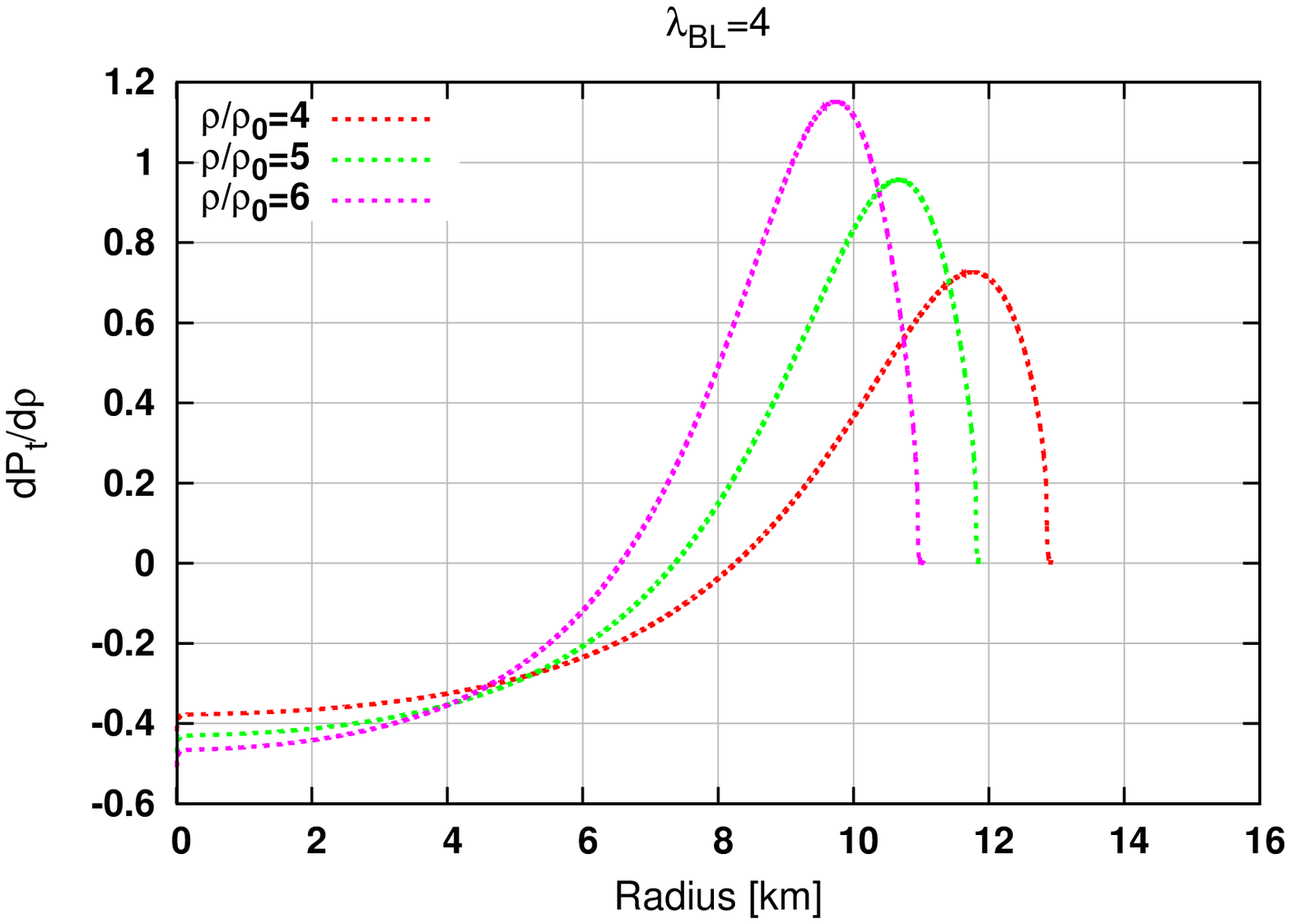}\\
\end{tabular}
\end{center}
\caption{Radial profile of transverse pressure is plotted for different values of central density using DDH$\delta$ EOS and $\lambda_{BL}=4$ (left panel). Corresponding transverse sound speed is plotted as a function of radius (right panel).}
\label{fig:ECO}
\end{figure*}

Positive anisotropy parameter yields higher compactness. It has been argued that by increasing the value of the anisotropic parameter sufficiently the black hole limit can be reached~\cite{Yagi2,yagi3}. But one should carefully examine how the transverse pressure behaves for those high anisotropic parameters. In the left panel of Fig. \ref{fig:ECO} transverse pressure is plotted as a function of radius using DDH$\delta$ and $\lambda_{BL}=4$ for different values of central energy density. We observe transverse pressure increases with the radius and finally near the surface it starts decreasing and vanishes at the surface. In order to obtain such behavior for the transverse pressure we need some exotic physical phenomenon that enhances the value of the transverse pressure. But in reality, we do not expect such type of physical processes. In the right panel of Fig. \ref{fig:ECO} radial profile of the transverse sound speed squared (defined as, $c_{s,t}^2=\frac{dp_t}{d\rho}$) is plotted for the same configuration used in left panel. The region where the transverse pressure increases with $r$, the transverse sound speed becomes negative. Also, for higher central density (here, $\frac{\rho}{\rho_0}=6$) transverse EOS becomes acausal. So, clearly, it is physically not possible to achieve the black hole limit by increasing the degree of anisotropy.

\end{document}